\documentclass[aps, pre,amsmath,amssymb,twocolumn,floatfix,showpacs]{revtex4} 
 \usepackage[usenames,dvipsnames]{color} 
 \usepackage[pdftex]{graphicx} 
 \usepackage{epsfig} 
 \input epsf 
 
\graphicspath{ {figs/} }

     {\begin{list}{}%
             {\setlength{\leftmargin}{#1}}%
             \item[]%
     }
     {\end{list}}

 \newcommand {\dr}{{\mathrm d}\mathbf{r}}

 \newcommand {\rr}{\mathbf{r}}

\begin{document} 
 
 \title{{Soft-core particles freezing to form a quasicrystal and a
crystal-liquid phase}}
 
 \author{A.J.~Archer$^\ast$, A.M.~Rucklidge$^\dag$ and E.~Knobloch$^\#$} 
 \affiliation{$^\ast$Department of Mathematical Sciences, Loughborough University, Loughborough LE11 3TU, UK\\ 
$^\dag$Department of Applied Mathematics, University of Leeds, Leeds LS2 9JT, UK\\ 
$^\#$Department of Physics, University of California at Berkeley, Berkeley, CA 94720, USA} 
 
 \begin{abstract} 
Systems of soft-core particles interacting via a two-scale potential are
studied. The potential is responsible for peaks in the structure factor of the
liquid state at two different but comparable length scales, and a similar
bimodal structure is evident in the dispersion relation. Dynamical density
functional theory in two dimensions is used
to identify two novel states of this system, the crystal-liquid state, in which
the majority of the particles are located on lattice sites but a minority
remains free and so behaves like a liquid, and a 12-fold quasicrystalline
state. Both are present even for deeply quenched liquids and are found in a
regime in which the liquid is unstable with respect to modulations on the
smaller scale only. As a result the system initially evolves towards a small
scale crystal state; this state is not a minimum of the free energy, however,
and so the system subsequently attempts to reorganize to generate the lower
energy larger scale crystals. This dynamical process generates a disordered
state with quasicrystalline domains, and takes place even when this large
scale is linearly stable, i.e., it is a nonlinear process. With controlled
initial conditions a perfect quasicrystal can form. The results are
corroborated using Brownian dynamics simulations.
 \end{abstract} 
\pacs{61.50.Ah, 61.44.Br, 05.20.-y, 64.70.D-} 
 \maketitle 
 \epsfclipon 
 
\section{Introduction}

In hard condensed matter systems, the structure of the crystalline states that
are formed is largely determined by the strength of the bonds between the atoms
or molecules in the system, the dependence of the bonds on the orientation of
the particles and the packing of the particles. In general, thermal
fluctuations and entropy are less important, unless one considers a system near
the melting transition. In contrast, entropy and temperature can be
all-important in determining the structure of soft matter systems.

For polymers in solution, the interactions between pairs of chains depend on a
delicate balance between energy and entropy \cite{Likos01}. When the
solvent is good, the polymer chains form an open structure and interactions
between pairs of polymers are largely repulsive and entropic in origin. On the
other hand, when the solvent is less good, the polymer exhibits a tendency to
collapse. In a good solvent, the strength of the repulsion depends on how
branched the polymer is. As a result the form of the effective interaction
between polymers can be tailored and controlled via the polymer architecture.
In star-polymers, for example, the effective interaction potential is
determined by the number of arms on each star \cite{Likos01, likos:prl:98, likos:harreis:02}.

The effective interaction between soft polymeric macromolecules is also soft.
Since the centres of mass need not coincide with any particular monomer, the
effective interaction potential between the centres of mass can actually be
finite for all values of the separation distance $r$ between the centres. In
this paper we discuss the structure, phase behavior and dynamics of a
two-dimensional (2D) model system of such soft core particles.

The model that we study consists of soft particles which have a `core' plus
`corona' (or shoulder) architecture. The particles interact via the following
pair potential:
 \begin{equation} 
 V(r)=\epsilon e^{-(r/R)^8}+\epsilon ae^{-(r/R_s)^8},
 \label{eq:pair_pot} 
 \end{equation} 
where $R$ is the diameter of the cores of the particles, and $R_s>R$ is the
diameter of the corona (or shoulder) of the particles. In addition to the two
length scales present in the potential, there are two energy scales: the energy
penalty for a pair of particle cores to overlap is $\epsilon(1+a)$ and the
energy penalty for just the coronas to overlap is $\epsilon a$, where $a$ is a
dimensionless parameter that determines the shoulder repulsion strength.

The particular form of the pair potential in Eq.~\eqref{eq:pair_pot} arises
from considering the effective interaction between the centres of mass of
certain dendrimers or star-polymers. For dendrimers, this potential applies if
the inner generations of monomers are of one kind (hydrophobic, say), while the
outer generations are of another kind (hydrophilic, say). Similarly, if a star
polymer is made of diblock copolymers, then with a suitable choice of the block
length ratio, the effective interaction potential between the centres of mass
is expected to be of the form in Eq.~\eqref{eq:pair_pot}~\cite{MKL08, LBLM12}.

\begin{figure}
 \includegraphics[width=0.95\columnwidth]{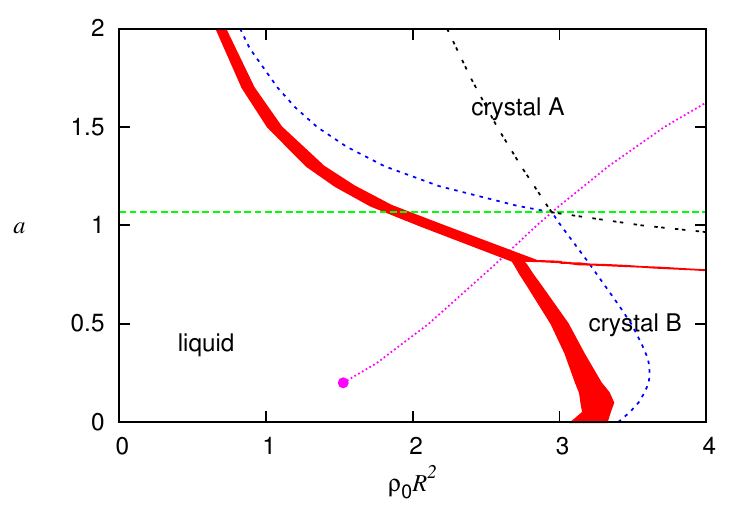} 
 \caption{{(Color online)} Phase diagram in the $(\rho_0,a)$ plane, where $\rho_0$ is the
average density. The system exhibits three phases: a liquid and two crystalline
phases. The `crystal A' is hexagonal, with a large lattice spacing, while
`crystal B', which is also hexagonal, has a smaller lattice spacing. The
small-dashed blue line is the linear instability threshold (spinodal) for the
uniform liquid. The large-dashed green horizontal line is where the two
principal peaks in the static structure factor $S(k)$ have the same height, while the
pink dotted line terminating in a circle is the locus where the two peaks in
the dispersion relation $\omega(k)$ have equal height. The circle marks the
point where the smaller $k$ peak disappears.}
   \label{fig:phase_diag}
 \end{figure}

In Fig.~\ref{fig:phase_diag} we display the phase diagram for a system with
temperature $k_BT/\epsilon=1$ and $R_s/R=1.855$. The figure shows the
$(a,\rho_0)$ plane, where $\rho_0=\langle N\rangle/L^2$ and $\langle N\rangle$
is the average number of particles in area $L^2$. This phase diagram is
determined using density functional theory (DFT), which is described below. 
Ref.~\cite{ARK13} provides a brief account of some of the work elaborated 
here. At low densities $\rho_0$ the particles form a liquid state. However, as 
the density increases, the particles overlap and then freeze to form one of two
different crystalline states. The crystals are unusual: they are of the
so-called `cluster-crystal'
variety~\cite{LBLM12, MGKNL06, LMGK07, MGKNL07, LMMGK08, MCLFK08, ZCM10, WS14},
related to the fact
that the particles have a soft core. In the cluster-crystal multiple particles
occupy each lattice site. When the parameter $a=0$, the system reduces to
particles interacting via a simple soft potential with one length scale and one
energy scale. This is the generalized exponential model with exponent $n=8$, or
GEM-8 model fluid~\cite{MGKNL06, LMGK07, MGKNL07, LMMGK08, MCLFK08, ZCM10, WS14, MoLi07}.
In 2D, at the temperatures relevant here, this model exhibits just one
hexagonal crystal phase, with lattice spacing $\sim R$ that is approximately
constant with increasing density (note that at very low temperatures, a series of
isostructural phase transitions is expected \cite{ZCM10, WS14}).
Similarly, when $a\gg1$ the
contribution from the core of the potential becomes negligible and the fluid is again
approximately a GEM-8 system, but now the particles have the larger diameter
$R_s$ and a stronger repulsion energy. Thus, for large $a$, the system forms a
hexagonal crystal with lattice spacing $\sim R_s$. We henceforth refer to this
larger lattice spacing crystal as the `crystal A' phase, and the smaller
lattice spacing crystal as the `crystal B' phase.

When the difference $R_s-R$ is small, one can pass smoothly from one crystal
phase to the other as $a$ is varied (i.e., there is only one crystalline
phase). However, when the difference is larger, as in the case displayed
in Fig.~\ref{fig:phase_diag}, crystal A and crystal B are two distinct phases
separated by a phase transition at $a\sim {\cal O}(1)$
 \footnote{We estimate
that crystal A and crystal B are
distinct phases when $R_s/R\gtrsim1.6$, since the ratio $R_s/R$ needs
to be greater than this value for there to be a point on the linear instability
threshold line where there are two modes that are unstable (i.e.~the point in
Fig.~\ref{fig:phase_diag} where the dotted and dashed lines meet).}.
Indeed, many of the
interesting novel properties of the present model described below all
occur in the regime when $a\sim {\cal O}(1)$, because they stem from the
competition between the two different length scales, $R$ and $R_s$, and the two
different energy scales, $(1+a)\epsilon$ and $a\epsilon$.

Two of the most striking properties of this system are: (i) When quenched to
certain regions of the phase diagram, where $a\sim {\cal O}(1)$, the system can
sometimes freeze to form states with quasicrystalline order. (ii) On examining
in detail the crystal A phase, again when $a\sim {\cal O}(1)$, we find that
there is a high proportion of mobile particles in the system, which is why we
refer to this phase as a `crystal-liquid'. It is crystalline, because the
majority of the particles in the system are frozen onto a regular hexagonal
lattice. However, a minority are `liquid', in the sense that they move
throughout the system. We find that the proportion of mobile particles in the
system can be as high as~7\%.

The quasicrystals (QCs) formed by the present system are a local equilibrium
state of the system, i.e., they are not the global minimum free energy state. 
They are found at state points in the portion of the phase diagram where the
thermodynamic equilibrium phase is the crystal-liquid A state. The QCs are 
formed by a particular dynamic mechanism when the system is quenched to certain 
regions of the phase diagram. 

In order to find parameter values at which QCs might be favored, we have
invoked understanding developed from analyzing mode interactions in the Faraday
wave experiment, in which a tray of liquid is subjected to vertical vibrations of
sufficient amplitude that standing waves form on the surface. This system
exhibits quasipatterns (the fluid dynamical analogue of quasicrystals), as 
discovered in the early
1990's~\cite{Christiansen1992, Edwards1993}, and two different mechanisms for
their formation have been identified (see~\cite{Rucklidge2009} for a more
detailed discussion). Briefly, patterns with $Q$-fold symmetry are
expressed as sums of modes with $Q$~wavevectors spaced at equal angles, and
weakly nonlinear theory is used to compute how waves with different
orientations affect each other. One mechanism relies on strong self-coupling to
downplay the effect of waves with different orientations~\cite{Muller1994,
Christiansen1995, Zhang1997}, so permitting
{8, 10, 12, 14, 16, 18, 20-fold or higher
quasipatterns \cite{Rucklidge2009}}. The second mechanism invokes nonlinear coupling between the
primary waves with secondary weakly damped (or weakly excited) waves, such that
primary waves with wavevectors separated by a certain angle determined
by the ratio of the primary to secondary wavenumber, are
favored~\cite{Edwards1993, Kudrolli1998, Arbell2002, Ding2006, Topaz2004,
Rucklidge2012, Lifshitz1997, Skeldon2015}. We invoke here this second
mechanism, as done in~\cite{Barkan2011, Barkan2014}, and select the
length scale ratio for our investigation to be $R_s/R=1.855$ 
(Sec.~\ref{sec:3}) in order that the ratio of the primary to secondary
wavenumbers is $2\cos(15^\circ)=1.932$, so favoring dodecagonal quasicrystals.

In fact the mechanism for QC formation that we actually observe differs from 
either of the two mechanisms described above. The QCs form when the uniform liquid is linearly unstable
against density fluctuations with a small wavelength that is close to that of
the lattice spacing of the crystal B phase but stable with respect to wavelengths
comparable to that of crystal A. Thus, in the initial stages after a
quench the system appears to be forming the crystal B phase. However, the
minimum free energy structure is actually the larger lattice spacing crystal A
phase. In the subsequent {\it nonlinear} evolution, the system seeks to form
this larger lattice-spacing phase. However, being already patterned with the
shorter length scale from the early stage linear dynamics, the system cannot
always form a perfect crystal A and often forms a state with a mixture of
both the short and long length scales that sometimes turns out to have
quasicrystalline ordering, i.e., the Fourier transform of the density
distribution reveals the presence of 12-fold ordering. As one might expect from
such a mechanism, the structure that is formed contains defects. However, by
carefully controlling the wave numbers of the density modulations prior to the
quench, the system can be induced to form a `perfect' quasicrystal.

Understanding the mechanisms by which soft matter QCs can form is becoming
increasingly important.
The possibility of designing soft-matter quasicrystals that self-assemble has
generated considerable interest at a fundamental level, leading to a burst
of experimental and theoretical
activity~\cite{Lee2010,Fischer2011,Iacovella2011,Xiao2012,Engel2015}.
Self-assembled soft-matter quasicrystals are of interest for a number of reasons,
not least because they
promise to provide a route to manufacturing materials and coatings with novel
optical or electronic properties arising as a consequence of their high degree
of rotation symmetry~\cite{Jin1999,Zoorob2000,Macia2012}.

{Although our focus is on polymeric soft matter QCs, we should also mention that there are other (colloidal) soft matter systems that form QCs
\cite{Dzugutov, denton1998stability, RothDenton2000, KeysGlotzer07, engel2014computational, dotera2014mosaic}. These also have pair potentials
involving more than one length scale, but owing to the particles having a
hard core, the local structure of these materials differs from that described below, as does the resulting phase behavior.}

This paper is structured as follows: In Sec.\ \ref{sec:theory} we describe the
DFT and dynamical DFT that we use to determine the structures formed by the
system. In Sec.\ \ref{sec:3} we discuss the properties of the uniform liquid state,
presenting results for the radial distribution function $g(r)$, the static structure factor
$S(k)$ and the dispersion relation $\omega(k)$. We then present results relating to
the solid states that are formed, focusing on the crystal-liquid state in
Sec.\ \ref{sec:4} and on QC formation in Sec.\ \ref{sec:QC}. This section includes a 
discussion of our numerical results and their relation to other mechanisms of QC formation 
from the literature that are relevant to soft-matter systems. The paper concludes in 
Sec.\ \ref{sec:conc} with a few concluding remarks.

 \section{Theory for the system}
 \label{sec:theory}

We use DFT \cite{Evans79,Evans92,Lutsko10,HM} to determine the structure,
thermodynamics and phase behavior of the system. To describe the dynamics of
the system when it is out of equilibrium, we use dynamical density functional
theory (DDFT) \cite{MaTa99,MaTa00,ArEv04,ArRa04}. The thermodynamic grand
potential of the system is a functional of the one-body density distribution
$\rho(\rr)$ of the particles:
 \begin{equation}\label{eq:grand_pot}
 \Omega[\rho(\rr)]=F[\rho(\rr)]+\int \dr \rho(\rr)(\Phi(\rr)-\mu),
 \end{equation}
where $\mu$ is the chemical potential, $\Phi(\rr)$ is the external potential
and $F[\rho]$ is the intrinsic Helmholtz free energy of the system, which is
composed of two contributions:
 \begin{equation}\label{eq:F}
 F[\rho(\rr)]=k_BT\int \dr \rho(\rr)\left[\ln(\rho(\rr)\Lambda^2)-1\right]+F_{ex}[\rho(\rr)].
 \end{equation}
The first term is the ideal-gas contribution, with $k_B$ the Boltzmann
constant, $T$ the temperature and $\Lambda$ the thermal de Broglie wavelength.
The second term, $F_{ex}$, is the excess (beyond ideal gas) portion describing
the contribution to the free energy stemming from the interactions among the
particles. The equilibrium density profile of the system at a given state point
$(\mu,T)$ is that which minimizes $\Omega[\rho]$, i.e., which satisfies the
equation
 \begin{equation}\label{eq:min_eq}
 \frac{\delta \Omega[\rho(\rr)]}{\delta \rho(\rr)}=0.
 \end{equation}
For systems of soft-core particles such as those we consider here, the
following rather simple mean-field approximation is remarkably accurate
\cite{Likos01, LLWL00, LBH00}:
 \begin{equation}\label{eq:F_ex}
 F_{ex}[\rho(\rr)]=\frac{1}{2}\int \dr \int \dr' \rho(\rr)V(|\rr-\rr'|)\rho(\rr').
 \end{equation}
This functional generates the following simple random-phase
approximation (RPA) for the pair direct correlation function:
 \begin{equation}\label{eq:c_2}
 c^{(2)}(|\rr-\rr'|)\equiv-\beta\frac{\delta^2F_{ex}[\rho(\rr)]}{\delta\rho(\rr)\delta\rho(\rr')}=-\beta V(|\rr-\rr'|),
 \end{equation}
where $\beta=(k_BT)^{-1}$.

To calculate the density profile of the system at a given state point
$(\mu,T)$, we discretize the density profile on a square Cartesian grid and use
fast Fourier transforms to evaluate the convolution integrals in
$F_{ex}[\rho]$. We employ standard Picard iteration \cite{Roth10} to solve the
Euler-Lagrange equation obtained from Eqs.~\eqref{eq:grand_pot}--\eqref{eq:min_eq},
 \begin{equation}\label{eq:EL_eq}
 \ln[\rho(\rr)\Lambda^2]-c^{(1)}(\rr)+\beta\Phi(\rr)-\beta\mu=0,
 \end{equation}
where
 \begin{equation}\label{eq:c_1}
 c^{(1)}(\rr)\equiv-\beta\frac{\delta F_{ex}[\rho(\rr)]}{\delta\rho(\rr)}
 \end{equation}
is the one-body direct correlation function. For the RPA functional in
Eq.~\eqref{eq:F_ex} this gives
 \begin{equation}\label{eq:c_1_RPA}
 c^{(1)}(\rr)=-\int \dr' \beta V(|\rr-\rr'|)\rho(\rr').
 \end{equation}
Equation~\eqref{eq:EL_eq} can be rearranged to obtain the following expression for
the density profile,
 \begin{equation}\label{eq:EL_eq_2}
 \rho(\rr)=\rho_0\exp\left(-\beta\Phi(\rr)+c^{(1)}(\rr)-c^{(1)}[\rho_0]\right),
 \end{equation}
where $c^{(1)}[\rho_0]$ denotes the value of $c^{(1)}$ when
Eq.~\eqref{eq:c_1_RPA} is evaluated for the uniform density profile
$\rho(\rr)=\rho_0$. We note the result
$\rho_0=\Lambda^{-2}\exp(-\beta\mu+c^{(1)}[\rho_0])$ showing that the average
density in the system $\rho_0$ is determined by the chemical potential $\mu$ or
vice versa.

Picard iteration of Eq.~\eqref{eq:EL_eq_2} corresponds to substituting the density
profile at step~$j$, $\rho^{(j)}(\rr)$, into the right side of Eq.~\eqref{eq:EL_eq_2} 
to obtain~$\rho_{\text{rhs}}^{(j)}(\rr)$. To
stabilize the iteration process these two density profiles at step $j$ are
mixed,
 \begin{equation}\label{eq:mix}
 \rho^{(j+1)}(\rr)=\alpha\rho_{\text{rhs}}^{(j)}(\rr)+(1-\alpha)\rho^{(j)}(\rr),
 \end{equation}
to obtain a new approximation for the density at step $j+1$. This equation is
then iterated until convergence is achieved. The value of the mixing parameter
$\alpha$ varies, depending on the state point and the type of density profile
to be calculated, but typically is in the range $0.001<\alpha\lesssim0.1$.

In the present study, we generate the initial guess for the density profile in several different ways. One choice is to use the density profile obtained from solving at a different state point. Another way is to start with the density profile $\rho(\rr)=\rho_0+\xi(\rr)$, where $\xi(\rr)$ is a small amplitude randomly fluctuating field. When the uniform fluid is linearly unstable (see below in Sec.~\ref{subsec:disp_rel}), this initial guess may converge to the density profile of the crystal. However, this method often results in density profiles containing defects. The chance of these forming is much less in smaller systems and so the density profile for a larger portion of a perfect crystal needs to be built up from the density profile obtained from a smaller system.

With this procedure the average density in the system
 \begin{equation}\label{eq:bar_rho}
 \bar{\rho}=\frac{1}{L^2}\int \dr \rho(\rr)
 \end{equation}
equals $\rho_0$ only when the system is in the uniform liquid state. For
the crystal, $\bar{\rho}\neq\rho_0$. This is because by iterating
\eqref{eq:EL_eq_2}, we actually select the value of the chemical potential
$\mu$. The density $\rho_0$ is the density of the uniform liquid for this value
of $\mu$, and $\bar{\rho}$ is therefore the density of the crystal
corresponding to this $\mu$ value. In calculations where we wish to specify the
average density to be $\rho_0$, we add an additional step to the Picard
iteration, where at each step $j$, after the mixing step given by
Eq.~\eqref{eq:mix}, we renormalize the density profile, whereby we replace
$\rho^{(j+1)}(\rr)$ with $f\rho^{(j+1)}(\rr)$, where
$f=\rho_0/[\frac{1}{L^2}\int \dr \rho^{(j+1)}(\rr)]$, cf.\
Eq.~\eqref{eq:bar_rho}. In all our discussions below, we do not distinguish
between $\bar{\rho}$ and $\rho_0$. We use $\rho_0$ to denote the average
density in all phases, but it should be borne in mind that when this refers
to a crystal phase, we mean the average density as defined in
Eq.~\eqref{eq:bar_rho}.

As presented, Picard iteration is simply a numerical algorithm for solving
the Euler-Lagrange equation and therefore for finding density profiles which
minimize the free energy. However, as shown in Sec.~\ref{sec:QC}, Picard iteration 
generates a series of density profiles that are often a fairly good approximation 
to the real dynamics as determined by DDFT (see Sec.~\ref{subsec:DDFT}), i.e., the 
index $j$ can be thought of as if it were proportional to the time $t$. In these 
cases we have used the fictitious dynamics generated by Picard iteration in place 
of the slower DDFT to survey the behavior at different points in the phase diagram.
 \begin{figure}[t]
 \includegraphics[width=0.7\columnwidth]{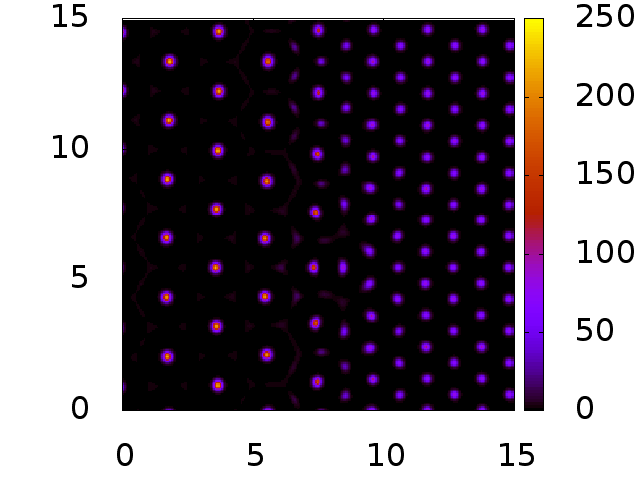}
 
 \includegraphics[width=0.7\columnwidth]{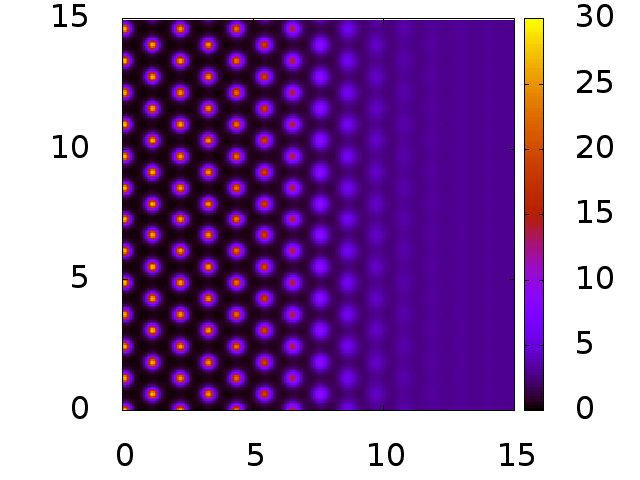} 
 \caption{{(Color online) Top: density profile in the $(x/R,y/R)$ plane} near an interface between coexisting crystal A and crystal B phases, for $a=0.75$. Because the lattice spacings of the two crystal structures are not commensurate defects along the interface are necessarily present. {Bottom}: density profile between the uniform liquid and the [1,1] interface of the crystal B phase, for $a=0.5$.}
   \label{fig:interfaces}
 \end{figure}

In Fig.~\ref{fig:interfaces} we display typical density profiles obtained from
DFT when the temperature $k_BT/\epsilon=1$ and $R_s/R=1.855$.
The phase diagram for this system is displayed in Fig.~\ref{fig:phase_diag}. In
the left panel of Fig.~\ref{fig:interfaces} we display the density profile at
an interface between coexisting crystal A and crystal B phases, for
$a=0.75$. Since the lattice spacings of the two crystal structures are
incommensurate, defects are necessarily present along the interface. In the
right panel of Fig.~\ref{fig:interfaces} we display the density profile across
an interface between the uniform density liquid on the right and the small
lattice spacing crystal B phase on the left, for $a=0.5$.

In the phase diagram displayed in Fig.~\ref{fig:phase_diag} the shaded (red
online) regions indicate coexistence between two different phases. The 
boundaries of these regions correspond to the densities of the two phases at
coexistence; recall that for two phases to coexist the chemical potential
$\mu$, the pressure $p\equiv -\Omega/L^2$ and the temperature $T$ must be equal in
the two phases. There is also a triple point, where all three phases coexist.
Note that in order to determine the minimum grand potential $\Omega$ one must also minimize
with respect to the computational domain size $L$. Actually, for hexagonal crystals one should
calculate the density on a domain of size $\sqrt{3}L\times L$. However, on
comparing results on such domains with those obtained on a square domain, we have confirmed
that as long as the (square) domain is sufficiently large that it contains many unit
cells, the slight strain energy contribution to the free energy is negligible.
{Most of the calculations presented here are for a system of size
$L=25.6R$ with periodic boundary conditions,
where the finite size effects for the regular crystal structures are negligible.
For QC structures, there are particular domain sizes (e.g., 8, 30 and 112 times the smaller scale) 
that allow for accurate
approximation to 12-fold QC structure \cite{Rucklidge2009}. Our results are for domains of 
size~30.
We remark further on this point in Sec.~\ref{sec:QC}.}

 \subsection{Liquid structure factor}

To characterize the structure of the liquid state, two quantities, the real
space radial distribution function $g(r)$ and the reciprocal space static
structure factor $S(k)$, are very useful \cite{HM}. The static structure factor
in the liquid phase is given by the relation
$S(k)=[1-\rho_0\hat{c}(k)]^{-1}$, where $\hat{c}(k)$ is the Fourier transform
of $c^{(2)}(r)$. Thus the RPA approximation for the structure factor in the
uniform liquid phase is
 \begin{equation}\label{eq:S_of_k}
 S(k)=\frac{1}{1+\rho_0\beta \hat{V}(k)}.
 \end{equation}
To determine the radial distribution function $g(r)$ one can insert the simple
RPA approximation \eqref{eq:c_2} into the Ornstein-Zernike equation \cite{HM}.
However, we choose instead to calculate $g(r)$ using the Percus test particle
method \cite{HM, AWTK14}. This gives a more accurate
approximation for $g(r)$ and also illustrates better the true accuracy of the
DFT that we use. The test particle method corresponds to fixing one of the
particles at the origin, so that $\Phi(\rr)=V(r)$ in Eq.~\eqref{eq:grand_pot},
and then calculating the density distribution of the remaining particles in the
presence of this fixed particle. The radial distribution function is obtained
from the resulting density profile via Eq.~\eqref{eq:min_eq}:
$g(r)=\rho(r)/\rho_0$. {In Ref.~\cite{AWTK14} results from this
RPA-test-particle theory were compared with the more sophisticated
hyper-netted-chain (HNC) theory for a very similar 2D soft-core system.
The agreement between the two is rather good, which gives us confidence
that the simple RPA DFT is accurate.}
We present typical results for $g(r)$ and $S(k)$ in Sec.\ \ref{sec:3} below.

 \subsection{Dynamics: time evolution of the density}
 \label{subsec:DDFT}

In addition to the equilibrium fluid structure, we also determine the
non-equilibrium fluid dynamics. Since we consider soft polymeric `blobs' in
solution, an appropriate approximation is to assume that the centres of mass of
the particles move via Brownian motion, i.e., via overdamped stochastic
equations of motion:
 \begin{equation}
 \dot{\rr}_i= -\Gamma\nabla U(\rr^N,t) + \Gamma{\bf X}(t),
 \label{eq:EOM}
 \end{equation}
where $i=1,..,N$ is an index that labels all the different particles in the
system, whose set of position coordinates we denote by
$\rr^N\equiv\{\rr_1,\rr_2, \cdots, \rr_N\}$. The mobility coefficient
$\Gamma=\beta D$, where $D$ is the diffusion coefficient, while ${\bf X}(t)$
denotes the random force on the particles due to the solvent thermal motion. We
assume in the standard way that ${\bf X}(t)$ is a Gaussian random variable
\cite{MaTa99,MaTa00,ArEv04,ArRa04}. The potential energy of the system is
 \begin{equation}
 U(\rr^N,t)=\sum_{i=1}^N\Phi(\rr_i)+\sum_{j>i}\sum_{i=1}^NV(|\rr_i-\rr_j|).
 \label{eq:pot_energy}
 \end{equation}
For a system of interacting particles with equations of motion given by
\eqref{eq:EOM}, we can use DDFT \cite{MaTa99,MaTa00,ArEv04,ArRa04} to determine
the time evolution of the fluid non-equilibrium density distribution
$\rho(\rr,t)$. In DDFT the dynamics is governed by
 \begin{equation}
 \frac{\partial\rho(\rr,t)}{\partial t} = 
  \Gamma \nabla\cdot\left[\rho(\rr,t)\nabla\frac{\delta\Omega[\rho(\rr,t)]}{\delta\rho(\rr,t)}\right],
 \label{eq:DDFT}
 \end{equation}
a result that follows on making the approximation that the non-equilibrium
fluid two-point density correlation function is the same as that in the equilibrium fluid with
the same one-body density distribution. Equation~\eqref{eq:DDFT} is thus an
approximation \cite{MaTa99,MaTa00,ArEv04,ArRa04}, but for soft-core fluids,
previous good agreement with the results from Brownian dynamics (BD) computer
simulations (i.e., from solving repeatedly Eqs.\ \eqref{eq:EOM} and then
averaging over the different realizations of the noise), gives us confidence
that Eq.~\eqref{eq:DDFT} provides a good approximation to the exact dynamics.

 \subsection{Dispersion relation}
 \label{subsec:disp_rel}
 
 An important quantity for understanding the behavior of the system is the
dispersion relation, $\omega(k)$. This relation determines the rate at which
density fluctuations in the uniform liquid grow ($\omega>0$) or decay ($\omega<0$) 
over time. Consider a uniform liquid with density $\rho_0$, with a superposed small amplitude
perturbation $\tilde{\rho}(\rr,t)\equiv\rho(\rr,t)-\rho_0$. Equation~\eqref{eq:DDFT}
shows that the perturbation evolves according to
 \begin{equation}\label{eq:DDFT_lin}
 \frac{\partial\tilde{\rho}}{\partial t}={\cal L}\tilde{\rho}+{\cal O}(\tilde{\rho}^2),
 \end{equation}
where ${\cal L}\equiv D\nabla^2-D\rho_0 \nabla^2c^{(2)} \otimes$ 
is a linear operator and $\otimes$ denotes a convolution, i.e., $c^{(2)}
\otimes\tilde{\rho}\equiv\int d{\bf r}' c^{(2)}({\bf
r}-{\bf r}')\tilde{\rho}({\bf r}')$. To obtain this result one must make a
functional Taylor expansion of $F_{ex}[\rho]$ \cite{ArEv04, ARTK12, AWTK14}.
Linearising Eq.~\eqref{eq:DDFT_lin} and decomposing $\tilde{\rho}$ into a sum
of different Fourier modes,
 \begin{equation}\label{eq:Fourier_modes}
 \tilde{\rho}({\bf r},t)=\sum_{\bf k}\hat{\rho}_{\bf k}e^{i{\bf k}\cdot{\bf r}+\omega(k)t},\qquad k\equiv|{\bf k}|,
 \end{equation}
leads to the dispersion relation \cite{ArEv04, ARTK12, AWTK14}
 \begin{equation}\label{eq:disp_rel}
 \omega(k)=-Dk^2[1-\rho_0\hat{c}(k)].
 \end{equation}
On combining this result with the RPA approximation \eqref{eq:c_2} we obtain
$\omega(k)=-Dk^2[1+\rho_0\beta\hat{V}(k)]$, a result closely connected to the
structure factor $S(k)$ defined in Eq.~\eqref{eq:S_of_k}. This connection
applies in the case of a stable uniform liquid.

The liquid state is described as being linearly stable if $\omega(k)<0$ for all 
wave numbers $k$ and linearly unstable when $\omega(k)>0$ for
some wave number $k$. This situation arises for state points deep inside the 
parameter regime where
the crystal is the equilibrium phase. The linear instability threshold is
defined as the locus in the phase diagram where $\frac{d
\omega(k)}{dk}\big|_{k=k_c}=0$ together with $\omega(k=k_c)=0$, i.e., the locus
where the maximum growth rate is zero. The location of this threshold is
displayed in Fig.~\ref{fig:phase_diag} as the blue short-dashed line.

 \section{Structure of the liquid}
 \label{sec:3}

 \begin{figure}
 \includegraphics[width=0.95\columnwidth]{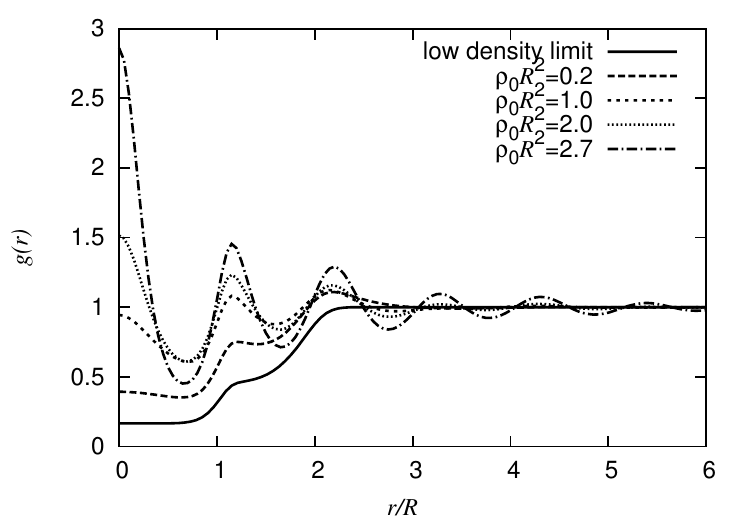} 
 \includegraphics[width=0.95\columnwidth]{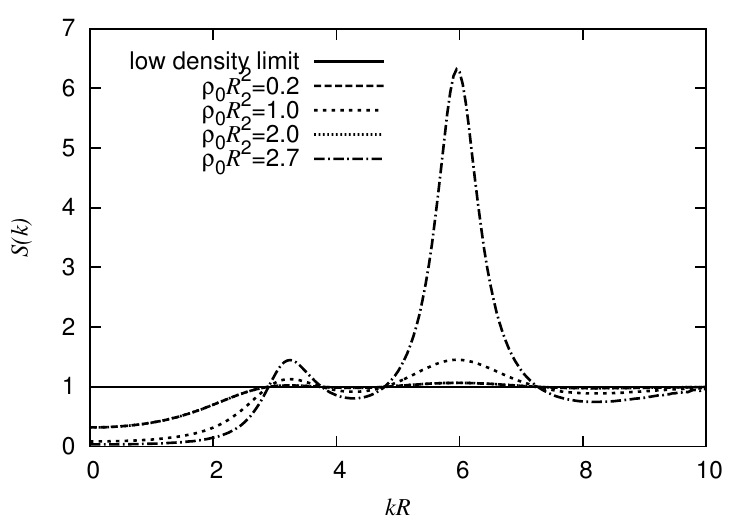} 
 \includegraphics[width=0.95\columnwidth]{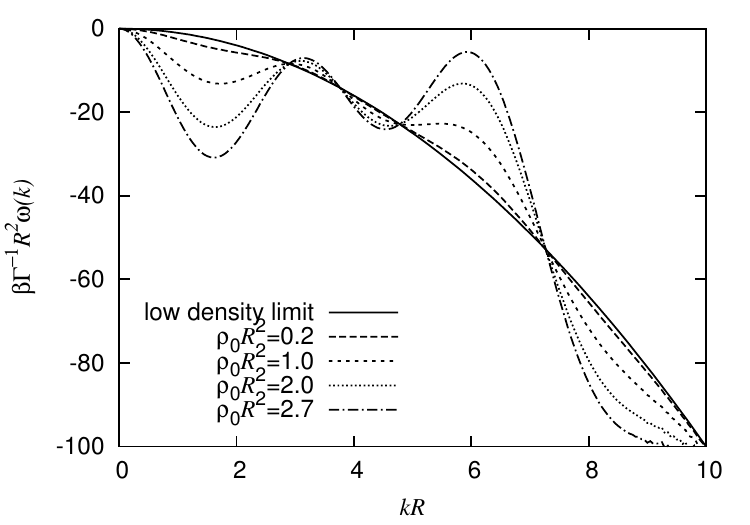} 
 \caption{Correlation functions characterising the liquid phase for increasing density $\rho_0$ as indicated in the key, for fixed $a=0.8$. Top panel: the radial distribution function $g(r)$; middle panel: the static structure factor $S(k)$; bottom panel: the dispersion relation $\omega(k)$.}
   \label{fig:correlations_a0_8}
 \end{figure}

 \begin{figure}
 \includegraphics[width=0.95\columnwidth]{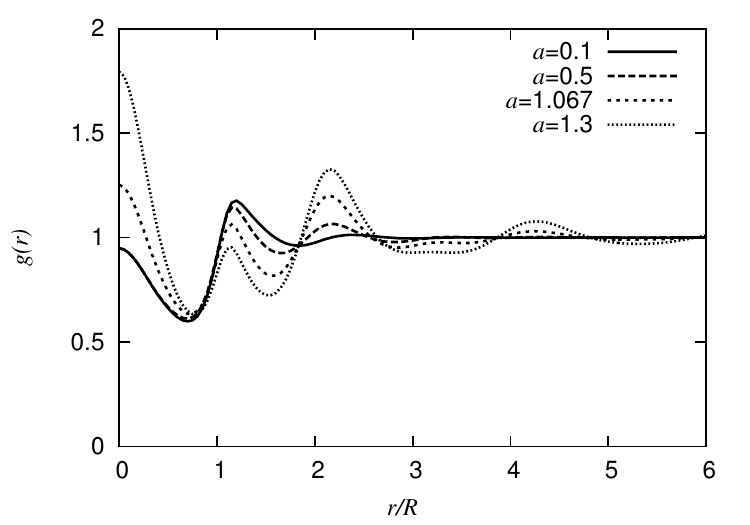} 
 \includegraphics[width=0.95\columnwidth]{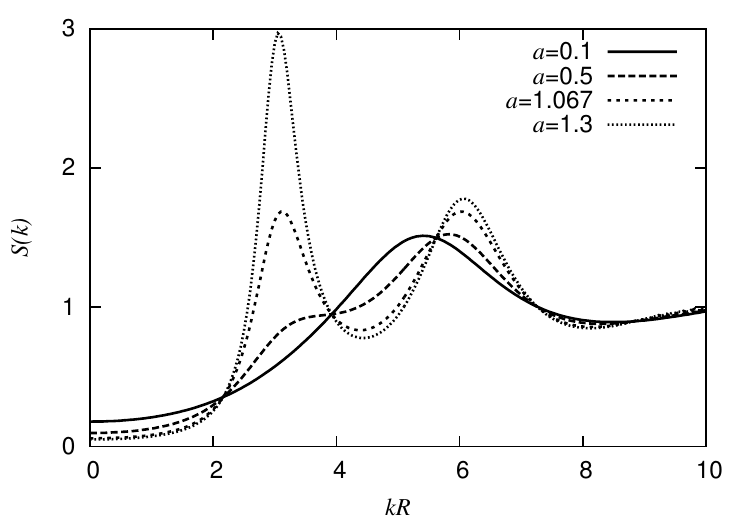} 
 \includegraphics[width=0.95\columnwidth]{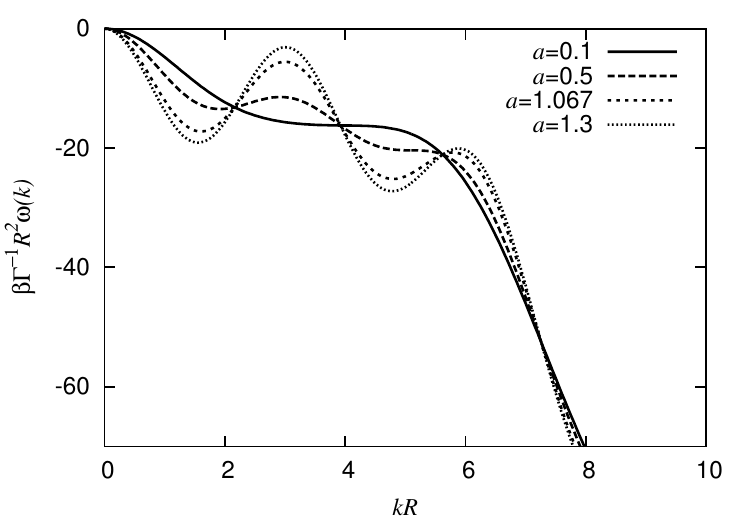} 
 \caption{Correlation functions characterising the liquid phase for a range of values of $a$ as indicated in the key, for fixed $\rho_0R^2=1.2$. Top panel: the radial distribution function $g(r)$; middle panel: the static structure factor $S(k)$; bottom panel: the dispersion relation $\omega(k)$.}
   \label{fig:correlations_rho1_2}
 \end{figure}

 \begin{figure}
 \includegraphics[width=0.95\columnwidth]{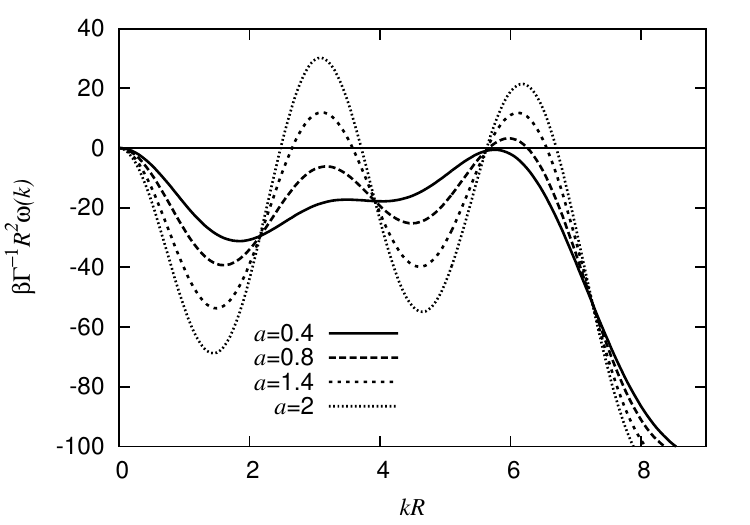} 
 \caption{The dispersion relation $\omega(k)$ for fixed $\rho_0 R^2=3.5$ and various values of $a$, as indicated in the key.}
   \label{fig:omega_rho3_5}
 \end{figure}
 
 \begin{figure} 
 \includegraphics[width=0.95\columnwidth]{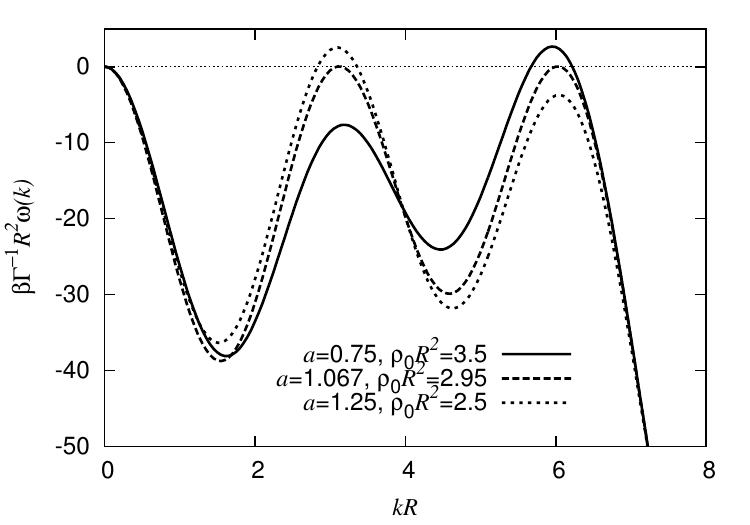} 
 \caption{The dispersion relation $\omega(k)$ at a series of points along a diagonal path in the phase diagram passing through the point at $\rho_0 R^2=2.95$ and $a=1.067$ at which the two modes $k_1R=3.12$ and $k_2R=6.03$ are simultaneously marginally unstable. Note that $k_2/k_1=1.932$.}
   \label{fig:omega_diagonal_path}
 \end{figure}

In this section we present some typical results for the radial distribution
function $g(r)$, the static structure factor $S(k)$ and the dispersion relation
$\omega(k)$ to illustrate the changes in the structure of the uniform liquid as
the average density $\rho_0$ and the shoulder height parameter $a$ are varied.

In Fig.~\ref{fig:correlations_a0_8} we display a series of results for fixed
$a=0.8$, as the density of the fluid is increased from zero to the value
$\rho_0R^2=2.7$, which for $a=0.8$ is the density of the liquid at coexistence
with the crystal B phase. In the top panel we display the radial distribution
function. In the limit $\rho_0\to0$ this is given by $g(r)=\exp[-\beta V(r)]$,
which exhibits a correlation hole for small $r$ due the particles seeking to
avoid overlaps. However, since the repulsion strength for full overlap at this
temperature is $\beta V(r=0)=\beta\epsilon(1+a)=1.8$, which is not that large,
$g(r\approx0)$ is positive, reflecting the fact that there is a nonzero
probability for particles to overlap completely, even at low densities. As the
density $\rho_0$ increases, the value of $g(r\approx0)$ also increases,
reflecting the fact that particles are forced to overlap more often.
Furthermore, oscillations develop in the tail of $g(r)$, at larger $r$. For
particles with a hard core, we would normally ascribe this behavior to packing
effects due to core exclusion. However, in the present system this is {
a largely energetic effect: as the density is increased, the overall energy is lowered if
some particles overlap with each other completely, thereby avoiding more
expensive partial overlap with many particles simultaneously, although the degree 
to which this occurs depends on the balance between energetic and entropic effects}.
This behavior is
also reflected in the fact that for $\rho_0R^2>1$, $g(r\approx0)>1$.
This value of $g(r\approx0)$ continues to increase as the density is increased.
For the case $\rho_0R^2=2.7$ we see that $g(r)$ is highly structured, with
a pronounced peak at $r=0$ indicating multiple overlaps. In fact, it is this
growing tendency to form clusters that drives the freezing into a
cluster-crystal when the density $\rho_0R^2>2.7$.

In the middle panel of Fig.~\ref{fig:correlations_a0_8} we display the
structure factor $S(k)$ obtained via Eq.~\eqref{eq:S_of_k} at the same density
values. We see that as the density is increased, $S(k)$ exhibits two peaks.
These reflect the correlations in the system with two characteristic length
scales, $R$ and $R_s$, and so the two peaks in $S(k)$ are (roughly) at the wave
numbers $\approx 2\pi/R_s$ and $\approx 2\pi/R$. The fact that the peak at
$\approx 2\pi/R$ is larger reflects the fact that for this value of $a$ the
particle core repulsions dominate the repulsions due to the
shoulder. As a result for this value of $a$ ($a=0.8$) the system freezes to
form the small lattice spacing crystal B phase.

In the lower panel of Fig.~\ref{fig:correlations_a0_8} we display the
dispersion relation $\omega(k)$ at the same series of state points. Except for
the limiting case of low density, $\omega(k)$ also exhibits two peaks, reflecting the
peaks in $S(k)$. For all the results displayed $\omega(k)\leq0$ for all $k$
from which we infer that the liquid is in fact linearly stable at these
densities. Indeed, at these densities the uniform liquid is the global minimum
free energy state. However, for $\rho_0R^2>2.7$, the global minimum corresponds
to that of the hexagonal crystal. As the density is further increased (not
displayed), the larger $k$ peak in $\omega(k)$ continues to grow in height, and when
$\rho_0R^2\approx3.2$ the peak growth rate $\omega(k=k_c)=0$, indicating that
the uniform liquid is now marginally unstable with respect to perturbations
with wave number $k_c\approx2\pi/R$. The resulting linear instability threshold
(spinodal) line is displayed as the blue short-dashed line in
Fig.~\ref{fig:phase_diag}.

In Fig.~\ref{fig:correlations_rho1_2} we display $g(r)$ (top), $S(k)$ (middle)
and $\omega(k)$ (bottom) at fixed density $\rho_0 R^2=1.2$, as the shoulder
height parameter $a$ is varied. For small values of $a$, we see that $g(r)$
exhibits a peak just beyond $r=R$, since this is the effective diameter of the
particles. However, as $a$ increases, increasing the shoulder height, this peak
decreases in height while another peak develops just beyond $r=R_s$,
reflecting the growing dominance of the shoulder in determining the
correlations in the liquid. The liquid with density $\rho_0R^2=1.2$ and $a=1.3$
is at phase coexistence with the crystal A phase. The behavior observed in
$g(r)$ is, of course, reflected in the structure factor shown in the middle
panel of Fig.~\ref{fig:correlations_rho1_2}. Specifically, for small $a$ there is a single
peak in $S(k)$, at $kR\approx5.5$. As the shoulder height $a$ increases, this
peak moves slightly towards larger $k$, and a second peak develops at
$kR\approx3$, i.e., at a value of $k$ that is a little below the value
$2\pi/R_s$. The latter reflects the growing importance of the length scale
$R_s$ in the particle correlations in the liquid. As $a$ increases further the
peak at smaller $k$ overtakes the larger $k$ peak. The two peaks in $S(k)$ have
equal height at $a=1.067$, irrespective of the fluid density. The locus of this
point is displayed as the green long-dashed line in Fig.~\ref{fig:phase_diag}.
The lower panel of Fig.~\ref{fig:correlations_rho1_2} which displays the
dispersion relation $\omega(k)$ also shows the development and growth of a peak
at $kR\approx3$. Increasing $a$ beyond the values displayed in this figure
shows that this peak continues to grow in height until $\omega(k)>0$ for
$kR\approx3$, indicating the uniform fluid becomes linearly unstable. In
Fig.~\ref{fig:phase_diag} we display the locus along which the two principal
peaks in $\omega(k)$ are of equal height using a pink dotted line. Along this
line the growth/decay rates for density fluctuations with these two wave
numbers are the same.

In Fig.~\ref{fig:omega_rho3_5} we display the dispersion relation $\omega(k)$
for fixed $\rho_0 R^2=3.5$ and various values of $a$. For the case $a=0.4$
there is one main peak in $\omega(k)$, with its maximum close to zero,
indicating that this state point is close to but slightly outside the linear
instability threshold. As $a$ increases this peak grows in height and also
shifts to slightly larger wave numbers, as the uniform liquid becomes
linearly unstable. At the same time a second peak starts to develop at $kR\approx3$ and
becomes the dominant peak for $a>1.4$. Since $\omega(k)$ determines the growth
rate of density fluctuations in the unstable liquid, the figure reveals a
transition between the fastest growing modes at small $a$ to those at large
$a$; this transition takes place along the pink dotted line in
Fig.~\ref{fig:phase_diag}.

This can also be seen in Fig.~\ref{fig:omega_diagonal_path}, which displays
the dispersion relation along a diagonal path in the phase diagram passing through
the point $(\rho_0 R^2,a)=(2.95,1.067)$, corresponding to the cusp in the blue 
short-dashed marginal stability threshold line in Fig.\ \ref{fig:phase_diag}.
At this point two modes with wavevector ratio $k_2/k_1=1.932$ are marginally
unstable.

 \section{The crystal-liquid state}
\label{sec:4}

 \begin{figure}
   \centering
   \includegraphics[width=0.7\columnwidth]{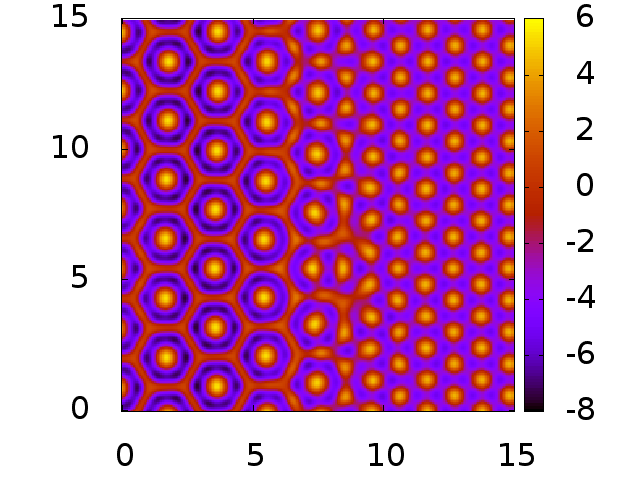} 
   \caption{{(Color online)} The density profile in the left panel of Fig.~\ref{fig:interfaces} displayed in terms of the logarithm of the density, $\ln[\rho(\rr)R^2]$, {plotted in the $(x/R,y/R)$ plane}. This representation allows one to see the fine structure of the density profile away from the principal peaks. Note in particular the honey-comb structure surrounding each of the peaks in the crystal A phase on the left of the interface.}
   \label{fig:log_rho_interface}
 \end{figure}

 \begin{figure}
   \centering
   \includegraphics[width=0.49\columnwidth]{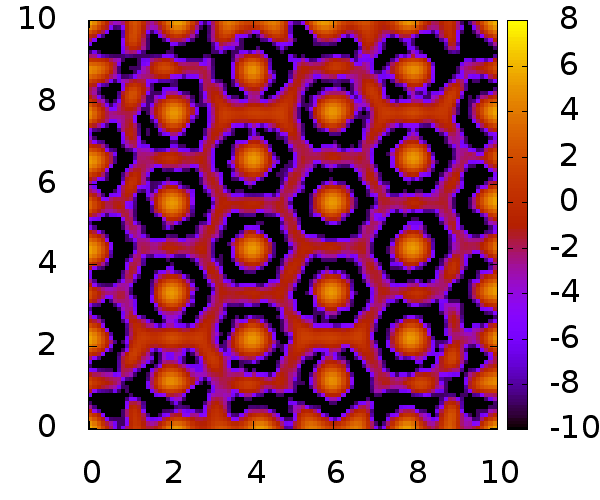}
   \includegraphics[width=0.49\columnwidth]{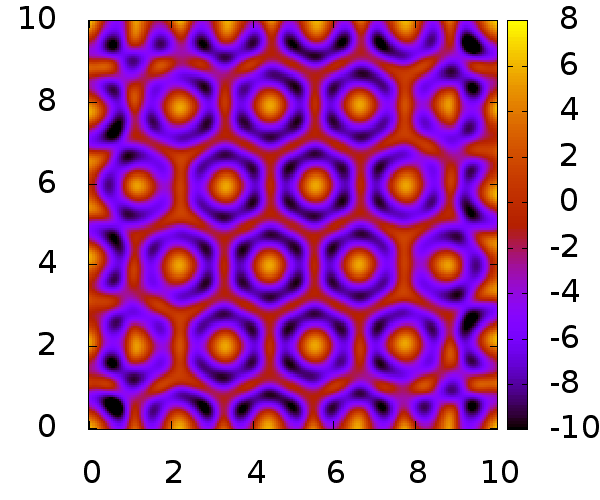} 
   
   \includegraphics[width=0.7\columnwidth]{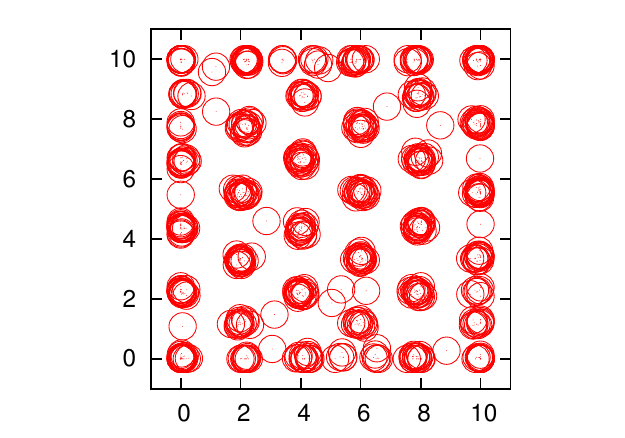} 
   \caption{{(Color online)} Top: $\ln [\rho(\rr)R^2]$ {in the $(x/R,y/R)$ plane} for a system of $N=600$ particles with $(a,\rho_0R^2)=(0.8,6)$ confined in a square region of side $L=10 R$ obtained from BD simulations (top left) and DFT (top right). The system forms crystal~A with a density profile consisting of an array of peaks surrounded by a connected network within which the particles are free to move -- this is the crystal-liquid state. Bottom: a snapshot from the BD simulation where each particle coordinate is plotted as an open circle.}
   \label{fig:snapshot}
 \end{figure}

 \begin{figure}
   \centering
   \includegraphics[width=0.48\columnwidth]{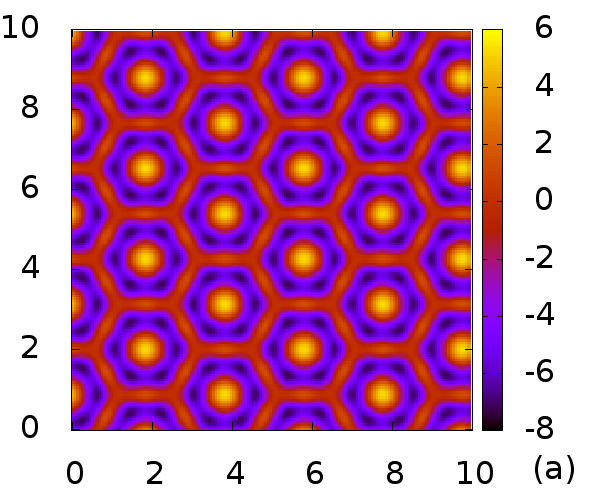} 
   \includegraphics[width=0.48\columnwidth]{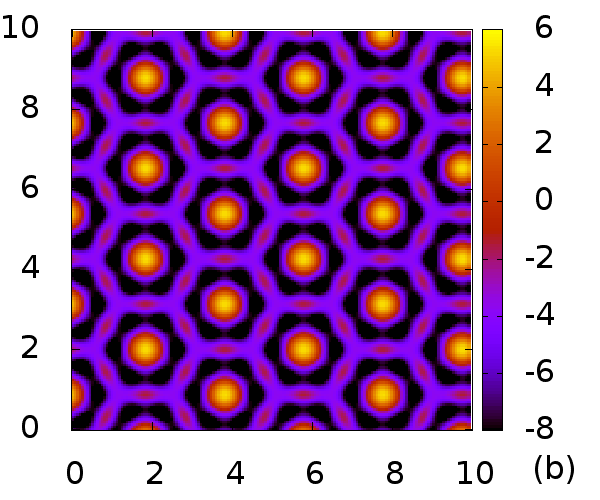} 
   
   \includegraphics[width=0.48\columnwidth]{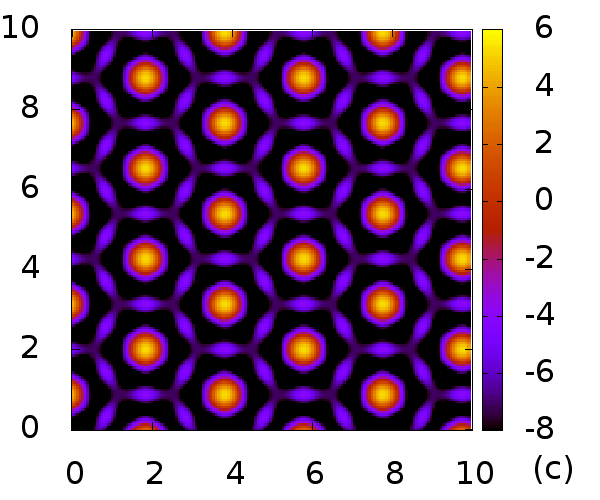} 
   \includegraphics[width=0.48\columnwidth]{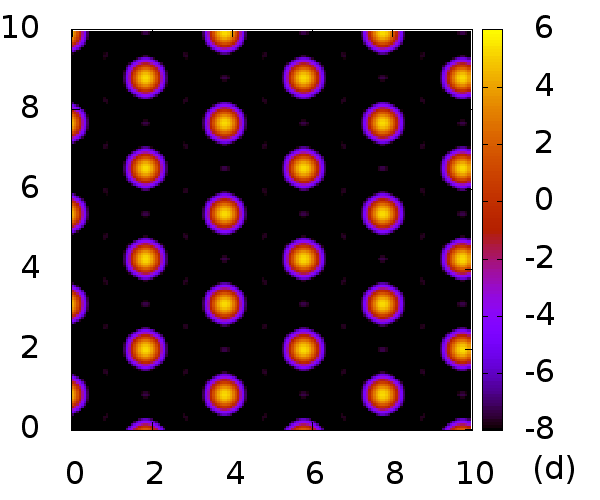} 
   
   \includegraphics[width=0.98\columnwidth]{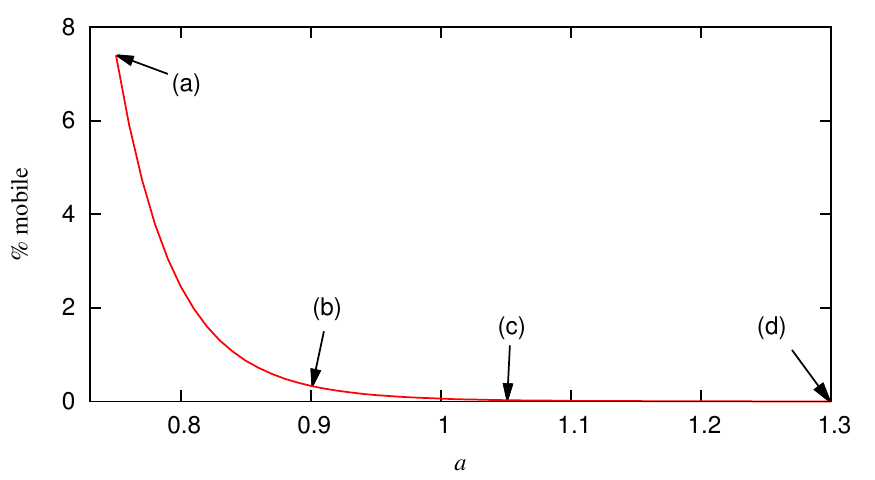}
   \caption{{(Color online)} (a)-(d) Plots of $\ln[\rho(\rr)R^2]$ in the crystal A phase {in the $(x/R,y/R)$ plane} for fixed $\beta\mu=39$ at the state points: (a) $(a,\rho_0R^2)=(0.75,4.1)$, (b) $(0.9,3.8)$, (c) $(1.05,3.5)$ and (d) $(1.3,3.1)$. The bottom figure shows a plot of the fraction of mobile particles that are in the liquid part of the density surrounding the density peaks. For $a<0.75$ crystal A is no longer the thermodynamic equilibrium crystal structure, and is replaced by crystal~B.}
   \label{fig:mobile_frac}
 \end{figure}

We now turn our attention to the density profiles in the crystal state. As
illustrated in Fig.~\ref{fig:interfaces}, at first sight the crystal A and
crystal B phases appear to be standard examples of hexagonally ordered
cluster-crystals. However, closer inspection of the density profile of the
crystal~A phase reveals that this is not the case, at least at state points
near to its coexistence with the crystal~B phase (i.e., at smaller $a$ values).
In Fig.~\ref{fig:log_rho_interface} we display the density profile in the
vicinity of the interface between the crystal~A and crystal~B phases shown in
Fig.~\ref{fig:interfaces}, but this time in terms of the logarithm of the
density, $\ln[\rho(\rr)R^2]$. This allows one to see the fine structure in the
density profile in the regions of space between the main peaks of the hexagonal
lattice. Here, we see an unbroken honeycomb-like network of density that
percolates throughout the crystal~A portion of the system, indicating that the
particles that contribute to this portion of the density profile are free to
move throughout the system. To confirm the existence of this striking
structure, we calculate, using both DFT and BD computer simulations, the density
profile for a system confined within a square confining potential $\Phi(\rr)$
with hard walls at $x=0,10R$, $y=0,10R$ so that $\Phi(x,y)=0$ for
$(0,0)<(x,y)<(10R,10R)$ and $\Phi(x,y)=\infty$ otherwise. The top left panel in
Fig.~\ref{fig:snapshot} shows the density profile obtained from the BD
simulations with $N=600$ particles and $\beta\epsilon=1$, $a=0.8$, i.e., the
average density in the box is $\rho_0R^2=6$. The BD result is obtained simply
by evolving in time the particles according to Eq.~\eqref{eq:EOM} and then
averaging over their positions to calculate the density profile. The top right
panel in Fig.~\ref{fig:snapshot} shows the corresponding density profile from
DFT. The remarkable agreement between the two confirms the validity of the DFT
approximation for this system. The bottom panel in Fig.~\ref{fig:snapshot}
shows a snapshot showing a typical configuration of the particles in the BD
simulation. The particle positions are indicated using open circles. Although
the majority of the particles are located on lattice sites, a significant
minority remain mobile, with the particles free to move in the density lanes
between lattice sites. The system thus consists of two dynamically distinct
populations. This is not observed in other pattern-forming 2D systems, such as
those in Refs.~\cite{Glaser_etal,ImRe04,ImRe06,Archer08}, where the dynamics of
all the particles are identical.

In Fig.~\ref{fig:mobile_frac}, bottom panel, we display the percentage of
mobile particles in crystal A as a function of the parameter $a$, for a fixed value of the
chemical potential, $\beta\mu=39$. This percentage is obtained by integration
over all portions of the density profile that are a distance $0.65R$ away from
the centre of the density peaks. Particles that contribute to this portion of
the density are defined to be mobile. Figures~\ref{fig:mobile_frac}(a)-(d)
display the logarithm of the density profile corresponding to the points
indicated in the lower panel. We see that as $a$ decreases the fraction of
mobile particles increases from zero, reaching a value of over~7\% at $a=0.75$.
We terminate the curve at this point because for $a<0.75$ crystal~A is no
longer the equilibrium crystal structure. It appears that as $a$ decreases
below this coexistence value, the growing proportion of mobile particles
triggers the formation of the smaller lattice spacing crystal~B phase, whereby
the mobile particles freeze to form the additional peaks of crystal~B.

 \section{The formation of {quasicrystals}}
 \label{sec:QC}

The role of resonant triads in the context of minimizing a free energy with one length scale 
has long been recognized~\cite{Alexander1978,Bak1985}. In particular in three dimensions
these triads can stabilize states with icosahedral symmetry. In two dimensions the presence
of two length scales implies the presence of two circles of wavevectors in Fourier space, 
and resonant triads involving wavevectors from these two circles can also contribute to 
stability of quasicrystals~\cite{Lifshitz1997}, provided the interaction coefficients are 
of the correct sign. With a radius ratio of the two circles equal to $2\cos(15^\circ)=1.932$, 
equilateral triads, $30^\circ$ triads and $150^\circ$ triads involving two vectors from one 
circle and one vector from the other increase the number of possible triads, and so the 
potential contribution to the free energy. This configuration leads to dodecagonal 
quasicrystals; with other radius ratios, the situation can be yet more complicated~\cite{Rucklidge2012}.
In fact, arguments based on the contribution to the free energy from
resonant triads only, important though these are, overlook the potential importance of 
higher order harmonics, whose coefficients may become arbitrarily large owing to the problem 
of small divisors that inevitably appears whenever quasiperiodicity and nonlinearity 
occur together~\cite{rucklidge2003convergence}. Thus a truncation of the theory at cubic 
order, a procedure widely used in the literature, remains to be properly justified, 
although Ref.~\cite{iooss2010existence} goes some way towards resolving the small
divisor issue.

\begin{figure*}

 \noindent
 \includegraphics[width=1.9in]{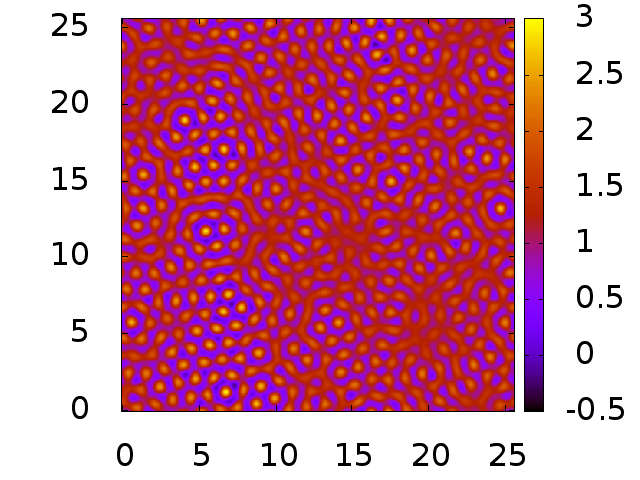} 
 \includegraphics[width=1.9in]{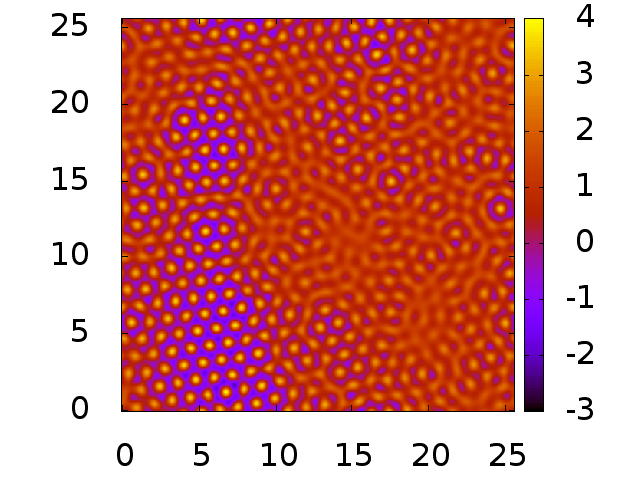}
 \includegraphics[width=1.9in]{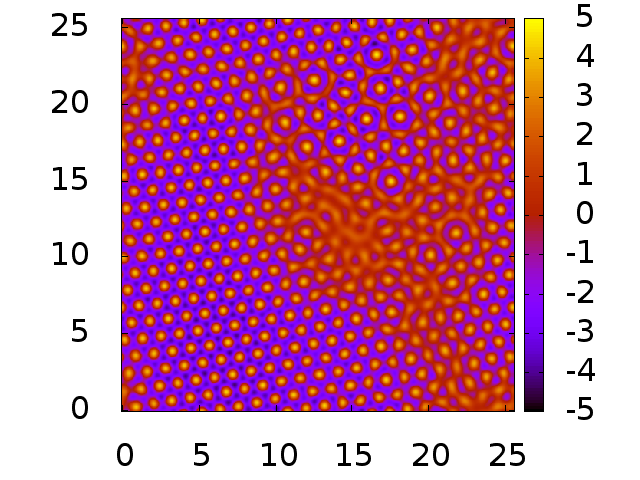} 

 \noindent
 \includegraphics[width=1.9in]{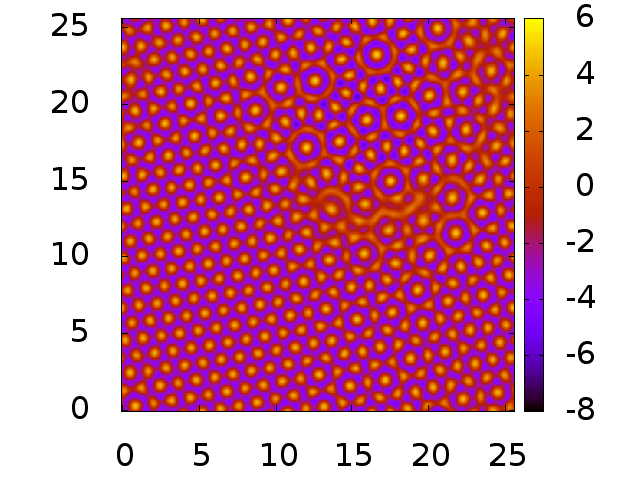} 
 \includegraphics[width=1.9in]{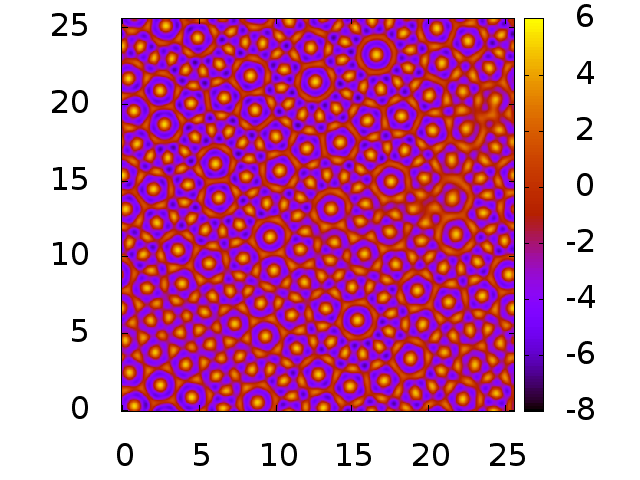} 
 \includegraphics[width=1.9in]{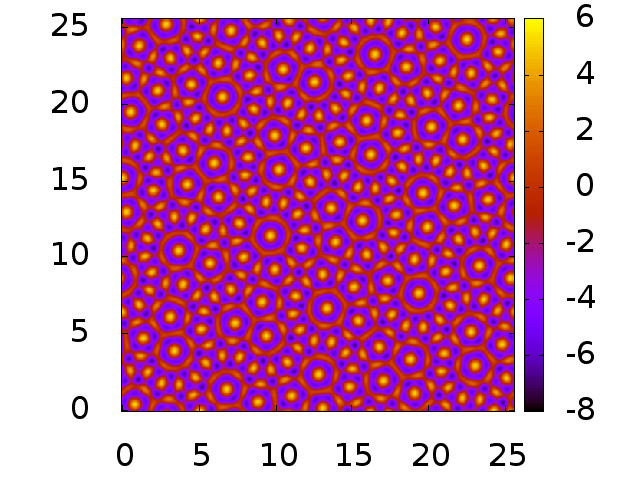} 
 \caption{{(Color online)} Snapshots of $\ln[\rho(\rr)R^2]$ {in the $(x/R,y/R)$ plane} obtained via Picard iteration for $a=0.8$ and $\rho_0 R^2=3.5$, revealing the evolution towards the equilibrium state for the same state point as the results displayed in the upper panel of Fig.~\ref{fig:B}. The dispersion relation at this state point is displayed in Fig.~\ref{fig:disp_rel1}(a). The panels along the top row, from left to right, correspond to times $t=30$, 32 and 35, and along the bottom row to $t=40$, 50, 200. Note that the system first forms the small length scale crystal (at time $t\approx30$). It then tries to form the longer length scale crystal. However, due to the small length scale already imprinted on the system, it cannot form a perfect large length scale crystal and ends up forming a disordered system with domains of QC ordering. The Picard iteration used to generate these figures does not locally conserve particle number (although it does conserve the total density in the system -- see Sec.~\ref{sec:theory}), but is much faster than the full DDFT and gives qualitatively similar results -- compare this figure with Fig.~\ref{fig:DDFT}, which is calculated with DDFT.}
   \label{fig:Picard}
 \end{figure*}

\begin{figure*}
 \noindent
 \includegraphics[width=1.9in]{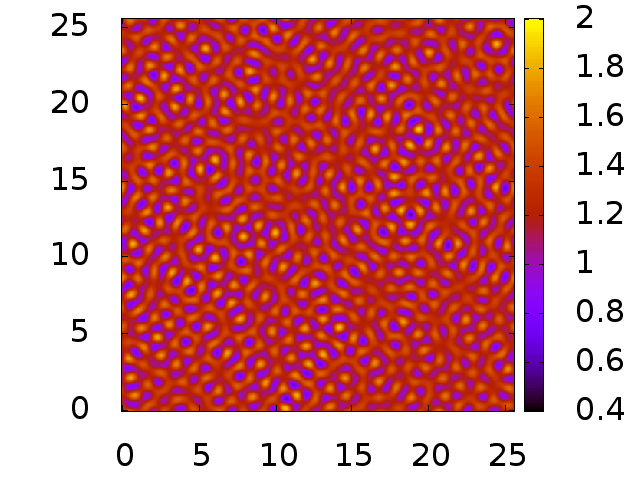} 
 \includegraphics[width=1.9in]{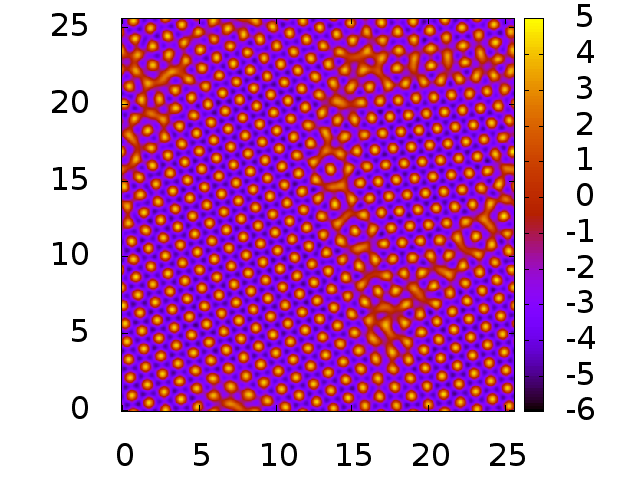} 
 \includegraphics[width=1.9in]{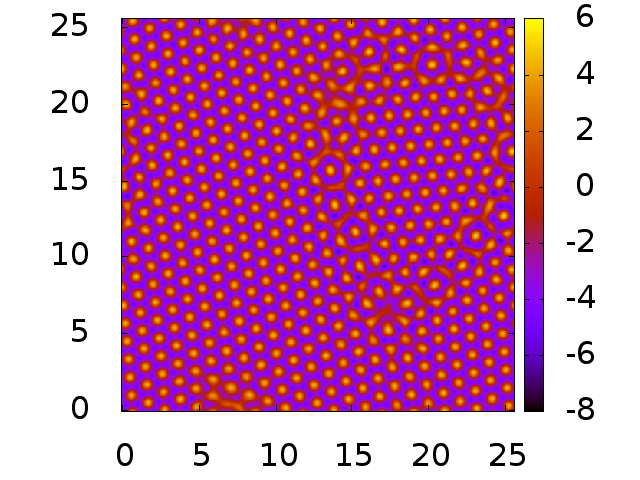} 

 \noindent
 \includegraphics[width=1.9in]{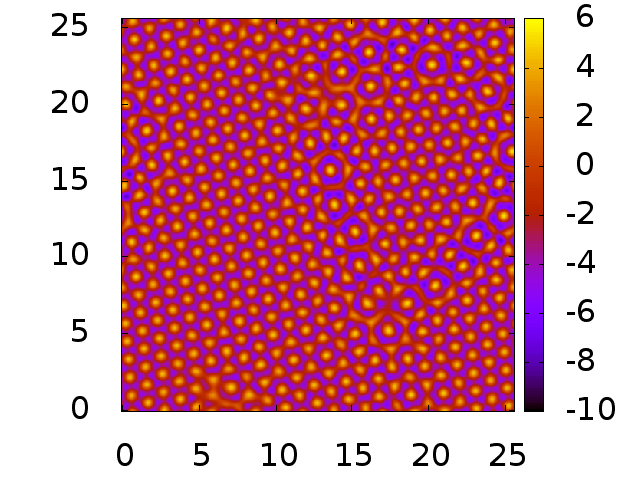} 
 \includegraphics[width=1.9in]{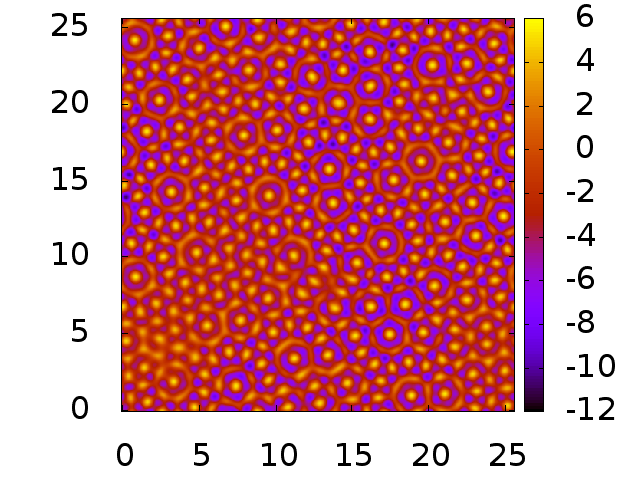} 
 \includegraphics[width=1.9in]{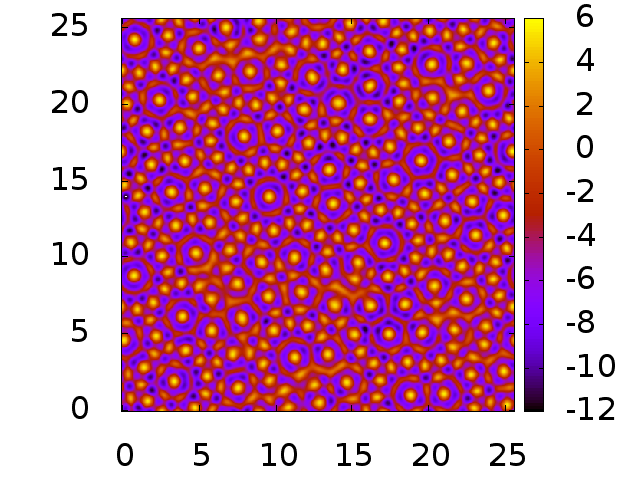} 
 \caption{{(Color online)} Snapshots of $\ln[\rho(\rr)R^2]$ {in the $(x/R,y/R)$ plane} obtained from DDFT, for $a=1.067$ and $\rho_0 R^2=3.5$. The dispersion relation at this state point is displayed in Fig.~\ref{fig:disp_rel1}(b). The panels along the top row, from left to right, correspond to times $t/\tau_B\equiv t^*=1$, 2 and 5, and along the bottom row to $t^*=10$, 20 and 40, where $\tau_B\equiv\beta R^2/\Gamma$ is the Brownian timescale. Note that the system first forms a small length scale crystal ($t^*=2$). It then tries to form the longer length scale crystal, initiated from a grain boundary -- see panels for $t^*=5$ and 10. However, because of the small length scale already imprinted on it, the system cannot form a perfect long length scale crystal and ends up forming a disordered system with domains of QC ordering -- see the final stationary profile at $t^*=40$.}
   \label{fig:DDFT}
 \end{figure*}

The mechanism identified below for stabilizing QCs in the present system also involves 
two length scales, but differs qualitatively from that just described (see also 
\cite{Barkan2011,Barkan2014}). In our case the system first forms the small length 
scale crystal phase. It is only when this phase is
almost fully formed (i.e., when the dynamics is far into the nonlinear regime)
that the longer length scale starts to appear, leading to the formation of the
QC (see Figs.~\ref{fig:Picard} and~\ref{fig:DDFT}). Thus, what we observe is in
fact a {hitherto unseen} mechanism for the formation of QCs.

The scenario for the formation of QCs
described in \cite{Barkan2011} requires the system (i) to be {{\em just}
inside the
linear instability line (i.e., $\rho$ is restricted to a small range beyond $\rho_\lambda$,
the value at the linear instability line),} 
and (ii) relies on the simultaneous {\it linear growth} of
two distinct wave numbers as in~\cite{Lifshitz1997}. In this scenario the role
of the higher order interactions (i.e., of nonlinearity) is to stabilize
the two length scale (QC) structures formed from the two linearly growing
scales \cite{Barkan2011}.

\begin{figure}
   \centering
   \includegraphics[width=3in]{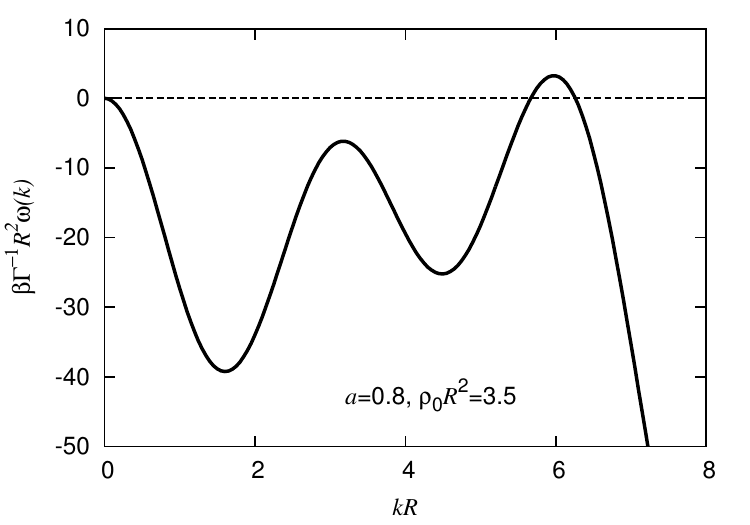} 
   
   \includegraphics[width=3in]{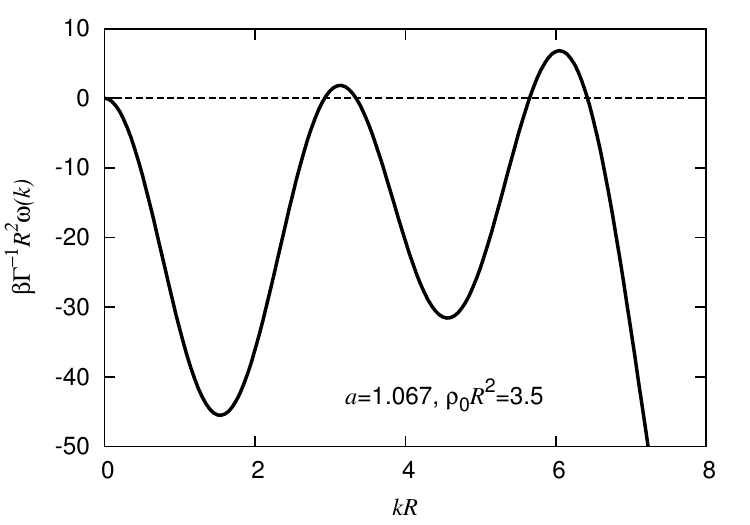} 
   \caption{Dispersion relation at the state point $a=0.8$ and $\rho_0 R^2=3.5$ (top, corresponding to Fig.~\ref{fig:Picard}) and $a=1.067$ and $\rho_0 R^2=3.5$ (bottom, corresponding to Fig.~\ref{fig:DDFT}). In both cases, QCs form at these state points. In the upper panel (Fig.~\ref{fig:Picard}) only one mode is unstable, corresponding to the smaller length scale crystal B. In the lower panel (Fig.~\ref{fig:DDFT}) two modes are unstable, but the growth rate for the smaller length scale crystal B is much larger than that for the larger length scale crystal A.}
   \label{fig:disp_rel1}
 \end{figure}

We contrast this scenario with that described here for a uniform liquid quenched 
to a region above the coexistence of the two crystal phases but below the pink 
dotted line in Fig.~\ref{fig:phase_diag}. In this regime the large $k$ peak 
dominates and small length scale density fluctuations grow rapidly 
(Fig.~\ref{fig:Picard}) as described by the dispersion relation 
in Fig.~\ref{fig:disp_rel1}(a). In this regime the system behaves as if it 
were going to form crystal B. However, the true minimum of the free energy
corresponds to the larger length scale crystal and this length scale is
linearly {\it stable} (Fig.~\ref{fig:disp_rel1}(a)). As a result, as the 
growing short-scale density
fluctuations reach the nonlinear regime, the system seeks to go to the longer
length scale structure but the smaller length scale imprinted from the linear
growth regime leads to frustration. We observe this type of behavior well away
from onset -- i.e., deep inside the linear-instability threshold, in contrast
to the scenario in \cite{Barkan2011}. In Fig.~\ref{fig:Picard} we display the 
resulting time evolution of the density profile as the system forms QCs. Since
only one mode is unstable the formation of the QCs that we find can only occur 
via the nonlinear mechanism described here.

In Fig.~\ref{fig:DDFT} we display DDFT results showing the formation of a QC 
structure at $a=1.067$ and $\rho_0R^2=3.5$ \footnote{Figure~2 of Ref.~\cite{ARK13} 
displays a QC formed at $(a,\rho_0 R^2)=(0.76,3.5)$.}. The dispersion relation 
corresponding to this 
state point is shown in Fig.~\ref{fig:disp_rel1}(b). We see that in this case 
the larger wavelength mode is no longer stable, although its growth rate is 
weak compared to that of the short wavelength mode. As a result
the system first forms the pure small length scale crystal (see, e.g., the
middle panel of the top row of Fig.~\ref{fig:DDFT} corresponding to
{$t^*=t/\tau_B=2$, where $\tau_B\equiv\beta R^2/\Gamma$ is the Brownian timescale}). 
However, over time, starting from a grain boundary, the system evolves a QC 
structure much as occurs at state point $a=0.8$, $\rho_0 R^2=3.5$. In both cases
this happens when the system is well away from the linear regime, in contrast to 
the weakly nonlinear QC mechanism proposed in Refs.~\cite{Barkan2011,Barkan2014}. 
Indeed, for $a=1.067$ the linear instability line is at $\rho_0 R^2=2.95$, implying 
that this state point corresponds, like $a=0.8$, $\rho_0 R^2=3.5$, to quite a deep quench. 
As a result, both snapshot series show that the linear growth regime introduces only {\it one} 
length scale, that of the small length scale crystal B phase -- despite the presence 
of the weakly unstable larger length scale in Fig.~\ref{fig:DDFT}. Figure~\ref{fig:DDFT} 
also confirms that the DDFT dynamics and the fictitious dynamics obtained from Picard 
iteration in Fig.~\ref{fig:Picard} are indeed qualitatively very similar. 

As explained above, our work shows that quasicrystalline structures can form
even when only {\it one} of the two scales introduced by our choice of the
potential is unstable; the instability forms nonlinear structures with this one
scale only but because these do not correspond to the global minimum of the
free energy which occurs at a distinct scale, the system attempts to shift the
structure to the thermodynamically preferred scale. This process leads to
frustration that is responsible for the formation of the QC state. This is a
qualitatively distinct mechanism of QC formation from that advocated in
Refs.~\cite{Barkan2011,Barkan2014} which requires that both scales are weakly unstable. 
As a result the latter theory is only capable of describing QCs that have very small 
amplitude. In contrast, our quasicrystalline states are present quite far from the onset
of instability and form from a periodic state via the nonlinear time-dependent process
described above.

 \begin{figure}[b]
 \includegraphics[width=0.53\columnwidth]{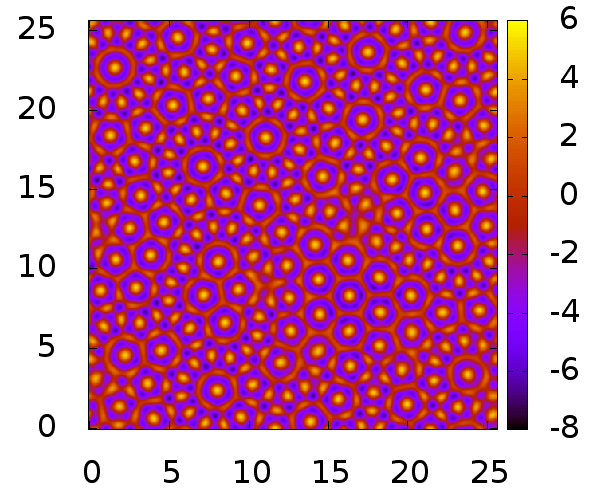} 
 \includegraphics[width=0.45\columnwidth]{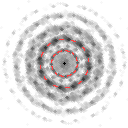} 
 
 \includegraphics[width=0.53\columnwidth]{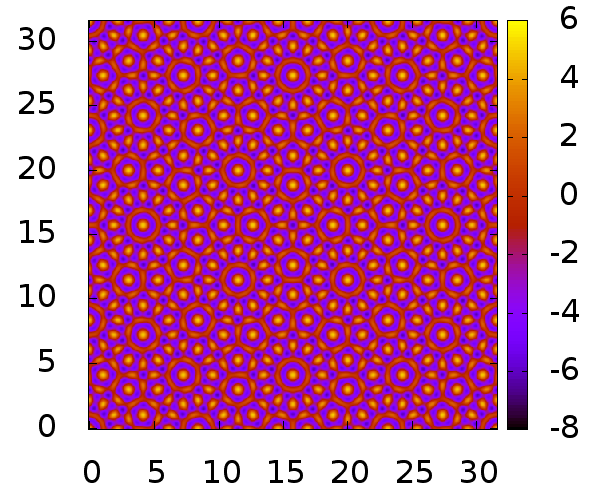} 
 \includegraphics[width=0.45\columnwidth]{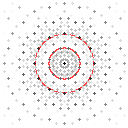} 
 \caption{{(Color online)} Left panels: plots of $\ln [\rho(\rr)R^2]$ {in the $(x/R,y/R)$ plane} obtained from DFT for $(a,\rho_0R^2)=(0.8,3.5)$. Right panels: the corresponding Fourier transforms. The latter exhibit 12-fold symmetry, which is indicative of QC ordering. The top density profile is obtained from random initial conditions, while the lower profile was formed starting from an initial density profile having QC symmetry.} 
 \label{fig:B} 
  \end{figure}
  
  \begin{figure}[t] 
     \centering
     \includegraphics[width=0.99\columnwidth]{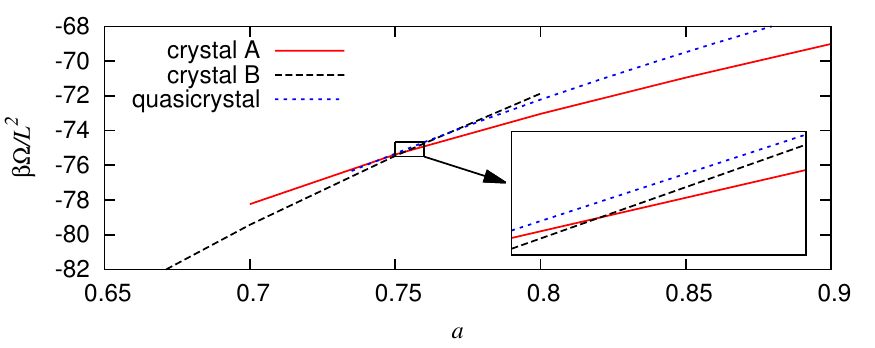} 
     \caption{{(Color online)} The grand potential density as a function of $a$ for fixed $\beta\mu=39$ for the two different crystal structures and also the QC solution displayed in Fig.~\ref{fig:B}. Near $a=0.75$ there is a point where all three have almost the same value of the grand potential, but the QC solution is never the global minimum (see inset). The crystal~A phase is of CL type throughout the range of $a$ shown.}
     \label{fig:free_energy}
  \end{figure}

  \begin{figure*}[t] 
     \centering
     \includegraphics[width=0.59\columnwidth]{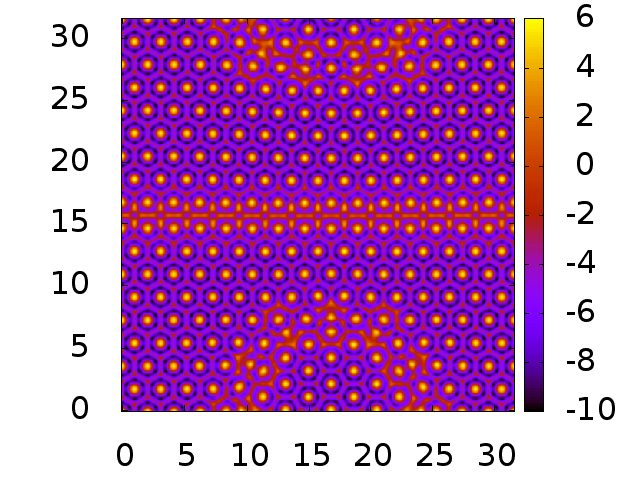}
     \includegraphics[width=0.59\columnwidth]{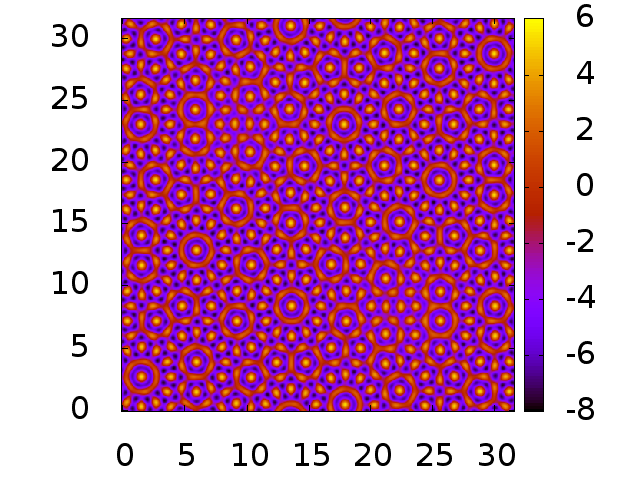}
     \includegraphics[width=0.59\columnwidth]{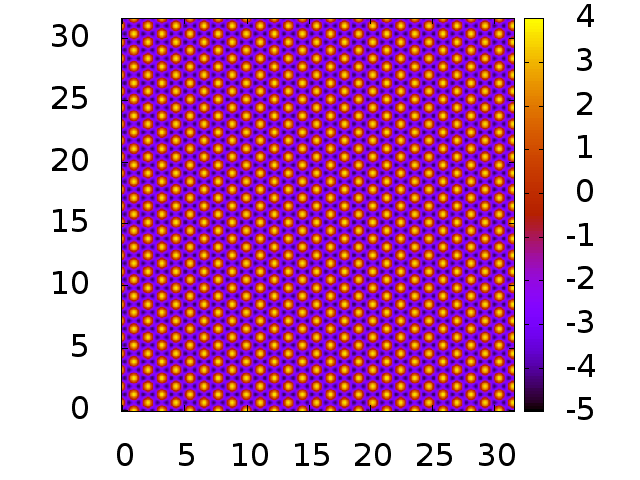}
     \caption{{(Color online)} Plots of $\ln[\rho(\rr)R^2]$ {in the $(x/R,y/R)$ plane} for density profiles obtained when starting from the `perfect' QC displayed bottom left of Fig.\ \ref{fig:B}, which has $R_s/R=1.855$, and then following the solution as $R_s$ is varied, for fixed $\mu$ and domain size. When $R_s$ is decreased, the QC remains stable until $R_s/R=1.77$, at which point the QC profile becomes linearly unstable and the Picard iteration then falls on to the crystal A profile (which contains defects), displayed left above. Alternatively, when $R_s$ is increased, the `perfect' QC solutions becomes unstable at $R_s/R=2.03$, where the iteration then switches to a different QC branch of solutions (middle above), before this finally becomes unstable at $R_s/R=2.19$, going to the crystal B brach of solutions displayed above right.}
     \label{fig:fall_off_QC}
  \end{figure*}

The calculation of the values of $a$ and $R_s/R$ used above to home in on the
parameter region where QCs might be observed was described in
Ref.~\cite{Barkan2011} as well as in earlier work \cite{Muller1994}. Once the
approximate parameter regime has been identified the details of what happens
depend on the values of $a$ and $R_s/R$. However, the QCs that we observe are
always {\it metastable} with respect to the periodic crystal. By this we mean
that they correspond to a local minimum of the free energy, but not to 
the global minimum (Fig.~\ref{fig:free_energy}). Thus the density profiles in 
Fig.~\ref{fig:B} are indeed
possible ground states (i.e., local minima), but not {\it the} ground
state (global minimum): the free energy of the state in the top panel in
Fig.~\ref{fig:B} is slightly higher than that of the lower panel, but both are
higher than that of the crystal A phase, which is the global minimum for this
state point. As $a$ increases beyond the range displayed in Fig.~\ref{fig:B},
the QC free energy increases more rapidly than that of the crystal A phase --
i.e., the trend revealed in Fig.~\ref{fig:free_energy} continues and the two
free energies do not approach one another again. In particular, the QC free
energy is far above that of the crystal A phase at $a=1.067$. This may also be so
for the QCs obtained in {Refs.~\cite{Barkan2011, achim, rottler}}.
In contrast, very recently
\cite{JTZS15} it has been shown that for the Lifshitz--Petrich free energy
\cite{Lifshitz1997}, QCs are indeed the global free energy minimum for certain
parameter values.

In Fig.~\ref{fig:B} we display both the QC density profiles and the
corresponding Fourier transforms. Both exhibit 12-fold ordering. In the upper
case, there is significant disorder in the system, which is not surprising
given the dynamical mechanism we observe for QC formation. However, it is
possible to facilitate a more ordered final state by choosing (for example) a
periodic domain $30$~times larger than the shorter of the
two lengthscales to allow for a circle of twelve vectors that are $29.98^\circ$
apart and whose lengths differ by $0.05\%$~\cite{Rucklidge2009}. Starting from an
initial condition with these twelve modes set to a small amplitude, we observe
that the system easily forms a `perfect' example of a QC (Fig.~\ref{fig:B}, 
lower panels).

In Fig.\ \ref{fig:fall_off_QC} we display density profiles obtained by taking this
`perfect' QC and then following the solution as the value of $R_s$ is changed.
We find that QCs remain linearly stable in the Picard iteration for $1.77<R_s/R<2.18$.
If $R_s$ is decreased to $R_s/R=1.77$, the QC solution at this point becomes
unstable and the Picard iteration leaves this solution and falls onto a crystal A
profile (which contains defects), displayed in the left hand panel of Fig.\ \ref{fig:fall_off_QC}.
If instead $R_s$ is increased, at $R_s/R=2.03$ the Picard iteration falls off the
branch of solutions corresponding to the `perfect' QC in bottom left of Fig.\ \ref{fig:B}
onto a different QC branch of solutions, which is displayed in the middle panel of
Fig.\ \ref{fig:fall_off_QC}. Further increasing $R_s$, this QC then becomes linearly
unstable at $R_s/R=2.18$ and the Picard iteration then goes to the crystal B
profile displayed in the right hand panel of Fig.\ \ref{fig:fall_off_QC}.

 \section{Concluding remarks}
 \label{sec:conc}

In this paper we have elaborated on the results of Ref.~\cite{ARK13} for a simple model soft core fluid that exhibits surprisingly rich phase behavior: two crystalline phases and a fluid phase. This stems from the fact that the pair potential between the particles has two length scales, $R$ and $R_s$, and two different energy scales, $a\epsilon$ and $(1+a)\epsilon$. The subtle balance of these leads to rich structuring and phase behavior. Pair potentials with these qualities arise as the effective interaction potentials between polymeric macromolecules. In particular, we believe that tailoring dendrimers with a `core' plus `shell' architecture should yield particles with effective interaction potentials akin to those considered here. Of course, the model system considered here is two-dimensional, so to observe the particular behavior reported here in an experimental system, the particles must be confined to an interface in order to create an effectively two-dimensional system. The natural next step to take after the work described here is to consider systems in three dimensions, where the phase behavior and the structures observed will be even richer. We are now embarking on work in this direction.

The two most striking aspects of the present model are: (i) The formation of the crystal-liquid phase, having two dynamically distinct populations of particles, some that are confined to the crystal lattice sites and others that are mobile, residing in a honeycomb-like network around the main density peaks. (ii) The formation of QCs. This aspect is particularly interesting, because the QC structures form via a mechanism that is distinct from any of the mechanisms that have been proposed previously. Namely, QC formation occurs following a deep quench of the uniform liquid to state points where it is unstable. At these state points, the global minimum of free energy corresponds to the large length scale crystal A phase. However, in the initial linear growth regime after the quench, a smaller length scale (corresponding to the small-length crystal B phase) grows the fastest, leading to the system becoming patterned with the ``wrong'' small-wavelength density modulations. When the system subsequently seeks to lower its free energy and hence to introduce the longer length scale, it remains ``stuck'' with some ordering on the small length scale. The final equilibrium structure generally consists of a mixture of the two length scales and may exhibit QC ordering, i.e., the Fourier transform may consist of a ring of 12 peaks. The resulting structure is in fact a local minimum of the free energy, but not the global minimum. The QCs formed via such a mechanism are, unsurprisingly, generally disordered, containing a mixture of domains with 12-fold ordering and domains of hexagonal ordering, corresponding to one or other of the two hexagonal crystal structures.

{The results presented here are for just one temperature. However, the important
quantities for determining the phase behavior of the model are the dimensionless quantities
$k_BT/\epsilon$, $a$ and $R_s/R$. Varying these determines the location in the phase diagram of the linear instability threshold, the point where both length scales are marginally
unstable and the ratio $k_2/k_1$. In the limit $a=0$, increasing $k_BT/\epsilon$ shifts
the linear instability threshold to higher density $\rho$ \cite{AWTK14}. Varying
the temperature by a modest amount should leave the behavior of the present system
qualitatively unchanged, merely shifting the regions where the crystalline phases occur to
higher densities.}

It is worth connecting the present work with related work {\cite{AWTK14,AWTK15,scheffler} on the freezing of binary mixtures of particles. The mixtures considered in Refs.\ \cite{AWTK14,AWTK15} also possess two length scales, owing to the fact that they are a binary mixture of soft particles of different sizes, and form multiple structures when a solidification front advances into an unstable uniform liquid. For a deep enough quench, such a front deposits behind it density modulations that are also of the ``wrong'' wavelength, thereby frustrating the formation of a well-ordered ``correct'' wavelength equilibrium crystal. In particular, the final equilibrium structures also contain a high degree of disorder}. In this case, the selection of the ``wrong'' wavelength is due to the dynamical nature of the length scale selection problem via an advancing front: the selected wavelength depends only on the linearization \eqref{eq:DDFT_lin}, whereas the global minimum free energy crystal structure is determined by the full DFT, which is highly nonlinear. This situation differs from the QC formation observed in the present work, yet there are similarities: both systems undergo a linear process that generates modulations with a length scale that does not correspond to the length scale of the equilibrium structure, which is determined by nonlinear processes. This naturally leads to an unanswered question: what happens when a solidification front advances in the present system? The front motion will generate a particular length scale, the linear growth of any local density modulations will produce a slightly different length scale while nonlinear interactions will seek to generate a third length scale. We anticipate that the interplay of such processes will inevitably lead to disordered structures.

\section*{Acknowledgements}
AJA thanks the Physics Department at UC Berkeley for kindly hosting him during the writing of this paper. The work of EK was supported in part by the National Science Foundation under Grant No. DMS-1211953.



 \end{document}